# Global Heliospheric Termination Shock Strength in the Solar-Interstellar Interaction

E. J. Zirnstein[1], R. Kumar[2], B. L. Shrestha[1], P. Swaczyna[1,3], M. A. Dayeh[4,5], J. Heerikhuisen[6], J. R. Szalay[1]

[1]Department of Astrophysical Sciences, Princeton University, Princeton, NJ 08544, USA
(ejz@princeton.edu)
[2]Princeton Plasma Physics Laboratory, Princeton, NJ 08543, USA
[3]Space Research Centre PAS (CBK PAN), Bartycka 18a, 00-716, Warsaw, Poland
[4]Southwest Research Institute, San Antonio, TX 78238, USA
[5]Department of Physics and Astronomy, University of Texas at San Antonio, San Antonio, TX 78249, USA
[6]Department of Mathematics and Statistics, University of Waikato, Hamilton, New Zealand

**A "heliospheric" termination shock (HTS) surrounds our solar system at approximately 100 astronomical units from the Sun, where the expanding solar wind (SW) is compressed and heated before encountering the interstellar medium. HTS-accelerated particles govern the pressure balance with the interstellar medium, but little is known about the HTS's global properties beyond in situ measurements from Voyager in only two directions of the sky. We fill this gap in knowledge with a novel and complex methodology: particle-in-cell, test particle, and MHD simulations, combined with a global minimization scheme to derive global HTS compression ratio sky maps. The methods utilize Interstellar Boundary Explorer observations of energetic neutral atoms produced from HTS-accelerated particles. Our results reveal unique, three-dimensional characteristics, such as higher compression near the poles during solar minimum, north-south asymmetries from the disparate polar coronal holes' evolution, and minimum compression near the flanks likely from SW slowing by mass-loading over a greater distance to the HTS.**

The interaction of the solar wind (SW) with the local interstellar medium (LISM) forms the heliosphere, a large structure that protects us from galactic cosmic rays[1,2] (though there is still a debate within the heliophysics community on the shape of the heliosphere, particularly the heliotail; see Kleimann et al.[3] and references therein). Beyond the critical Alfvén point, located at around 20 solar radii[4], the SW transitions from sub-Alfvénic to super-Alfvénic and the SW expands radially outward from the Sun at supersonic speeds. At two to three times the distance to Pluto, a shock (i.e., the "heliospheric termination shock", HTS) forms before the SW encounters the partially ionized interstellar medium. The outer boundary of the heliosphere is a three-dimensional surface called the heliopause, where the interstellar plasma outside the heliopause is slowed, compressed, and diverted around the heliosphere. Between the HTS and heliopause is the heliosheath (HS), containing a relatively hot plasma whose mean energy is primarily determined by the heating and acceleration of interstellar pickup ions (PUIs) at the HTS[5–8], in the HS[9–12], or most likely a combination of both. The location of the HTS[13–15] and heliopause[15–20], the plasma pressure in the HS[21–26], and how PUIs are accelerated at the HTS are all connected, forming a quasi-pressure-balance between the heliosphere and the VLISM that dynamically changes over time[27].

One of the key processes in this heliospheric pressure balance is the energization of interstellar PUIs at the HTS. There have been many theoretical and modeling studies on this topic, utilizing global magnetohydrodynamic (MHD), particle-in-cell (PIC), and test particle simulations[5,28–31,6,32,33]. The acceleration of PUIs depends on the strength and microstructure of the HTS, and fortunately the Voyager 2 spacecraft provided in situ observations of the HTS structure at scales smaller than the upstream advective PUI gyroradius[34,35]. Its measurements revealed a highly dynamic shock structure with PUI foot, ramp, overshoot, and undershoot. Voyager 2 crossed the HTS five times, implying a dynamically evolving shock with fast movements towards and away from the Sun. Each crossing showed slightly different kinetic-scale structures, but the larger scale structure of the HTS from the change in SW velocity revealed a large-scale compression ratio of approximately 2.5[35], and possibly higher[36] when including the energetic particle precursor[37], whereas the small-scale compression ratios during crossings TS-2 and TS-3 are $2.38 \pm 0.14$ and $1.58 \pm 0.71$, respectively[35], or an average of $1.98 \pm 0.36$. We have also learned more about PUI-mediated, quasi-perpendicular shocks from New Horizons' Solar Wind Around Pluto (SWAP) measurements of interstellar PUIs directly, revealing that the jump conditions at interplanetary shocks cannot be accurately determined from SW ions[38,39]; rather, PUIs are the only reliable source to derive the shock conditions (besides observing the magnetic field jump, which New Horizons is not equipped to measure). One of the strongest interplanetary shocks observed by SWAP occurred in October 2015, with a compression ratio between 2.5-3, with characteristics qualitatively similar to Voyager 2's observations of the HTS, but again showing little correlation between SW ions and the shock jump[40] (also see Shrestha et al.[41] for another observation of a strong shock). Voyager 2's PLS instrument[42], which only observed core SW particles at energies lower than those expected for PUIs[35] (even though PLS can observe particles up to 6 keV), observed lower compression ratios during TS-2 and TS-3.

Due to their preferential acceleration at the HTS, PUIs contain a significant fraction of the internal plasma pressure in the HS[43,21,24,25]. Energetic PUIs in the HS charge exchange with interstellar neutral atoms flowing into the heliosphere, creating energetic neutral atoms (ENAs) that propagate ballistically in all directions. Some ENAs make it to 1 au and are measured by the Interstellar Boundary Explorer (IBEX)[44–46]. The IBEX-Hi instrument measures ENAs in the 0.5-6 keV energy range, which approximately covers the range of energies that PUIs are believed to undergo "shock drift" acceleration at the HTS[6]. ENA spectra observed by IBEX-Hi therefore inform us of the post-accelerated PUI distribution in the HS and, using information from models of particle acceleration at the HTS and plasma flows in the HS, we may infer the HTS compression ratio across the sky. This will help answer one of the main scientific questions of the IBEX mission: "What is the global strength of the HTS?"[44], which is topical for both heliophysics and astrophysics studies of stellar-interstellar systems producing supersonic stellar winds[47,48]. As with our solar system, the shock strength will depend on the host star's stellar wind as well as the interplay between the interstellar dynamic and magnetic pressures.

**Results**

The goal of this study is to derive all-sky maps of the HTS compression ratio (i.e., its "strength"). No other study has attempted to construct a complete map of the HTS strength, only a few directions in the sky[49]; and those that have studied a few directions derived compression ratios using simplified theoretical equations to define the rate of heating of particles across the HTS as a function of shock strength, whereas our approach uses fundamental particle simulations.

We utilize IBEX-Hi observations in the spacecraft ram reference frame for the best statistics and make use of data from 2009-2016 to avoid the large increase in ENA fluxes that occurred starting in late 2016 due to a large increase in SW dynamic pressure. The latter is primarily to avoid situations where the globally distributed flux (GDF), or the primary signal of ENAs from the HS, changes abruptly (however, small contributions to the GDF also come from outside the heliopause[7,50]). We use IBEX-Hi observations from electrostatic analyzer (ESA) steps 3-6 (~1-6 keV) and exclude ESA 2 observations from our analysis, the reason for which is explained in the Methods section. We use IBEX ENA data transformed to proton fluxes in the HS plasma reference frame from Zirnstein et al.[51]. While in our previous study we developed a method to convert ENA fluxes in the spacecraft frame to proton fluxes in the simulated HS plasma frame, in the current study we use this new proton flux data from Zirnstein et al.[51] to compare IBEX data to simulations of test particle acceleration across the HTS and the resulting downstream distributions. Our global minimization scheme normalizes the simulated spectrum to the observations, thus not requiring us to know the thickness of the HS (see "Finding the Best-fit Compression Ratio at the HTS"). This then allows us to derive the global HTS compression ratio, which has never been done before and is an essential discovery for the science community. Several sky maps of the IBEX-derived proton fluxes in the line of sight-averaged HS plasma frame are shown in Figure 1.

**Upstream SW Conditions**

We require knowledge of the SW conditions upstream of the HTS to determine the downstream plasma properties corresponding to IBEX observation times (i.e., plasma conditions that created ENAs observed by IBEX). This requires propagating SW properties from 1 au to the HTS. In situ observations of SW parameters (e.g., density, temperature, speed, magnetic field) collected in the OMNI database from ACE and Wind measurements are used, which are observed within ±8° of the ecliptic plane. At higher latitudes, we utilize a model of SW speeds[52] derived from the latest release of interplanetary scintillation (IPS) observations[53]. As discussed in some previous studies[15,53,54], SW speeds derived from IPS at low latitudes tend to overestimate in situ observations in solar cycle 24. Therefore, a shift is applied to the IPS-derived speeds, separately for each year, down by a value at all latitudes to match OMNI at low latitudes[52]. Whether or not shifting the SW speed by the same value at all latitudes is correct cannot be determined at this stage[55]. Therefore, we include uncertainties in our analysis to account for this unexplained discrepancy (see Figure 24 in Porowski et al.[52]). This discrepancy may be due to a change in the relationship between density fluctuations and SW speed from solar cycle 23 to 24[53]; however, more in depth analysis may be required to understand discrepancies on a year-to-year basis and northern vs. southern hemispheres.

Because IPS observations only provide estimates of the SW speed, a reasonable assumption that the SW energy flux is independent of latitude is made[56], allowing us to use OMNI measurements of speed and densities to derive proton densities at high latitudes[57,54,52]. The plasma temperature and magnetic field magnitude are also extracted from the OMNI measurements at low latitudes and extrapolated to high latitudes using the following methods. For plasma temperature, we assume that at SW speeds of 750 km s$^{-1}$ the temperature is 250,000 K based on Ulysses measurements[58,59] and linearly interpolate from low latitude OMNI values to high latitudes with this reference point. Due to the variance observed in plasma temperature at high latitudes, we include an uncertainty of 10% at all latitudes. For magnetic field magnitude, we extract the field magnitude from OMNI measurements and apply the Parker spiral equations assuming that the transverse angle of the field at 1 au is 45°, yielding radial and tangential field components as a function of latitude. The alpha-to-proton density ratio is assumed to be latitude invariant; thus, we apply OMNI observations of the ratio at all latitudes. We note that this assumption does not significantly affect our results because the primary contribution to plasma pressure at the HTS is the SW proton dynamic pressure and thermal pressure from interstellar $H^+$ and $He^+$ PUIs[15].

SW parameters at 1 au as a function of latitude and time are propagated using a multi-fluid model to the HTS[15], providing the plasma conditions upstream of the HTS. Examples of the upstream SW speed for the two IBEX time periods of interest are shown in Figure 2. Note that the results are a function of IBEX ESA energy passband because the measurement time at 1 au may be the same, but the travel time for ENAs to get to IBEX from the outer heliosphere is different depending on their energy. The sharp transition in speed at longitude 180° is an effect of IBEX's annual map making process[44].

**Shock Micro-structure**

PIC simulations are used to determine the relationship between the upstream plasma properties, the shock compression ratio, and its micro-structure (i.e., shock width, overshoot). We ran a set of fully-kinetic, 2D PIC simulations[31] with different upstream Mach numbers, including SW protons, electrons, and interstellar $H^+$ PUIs, where the PUIs are represented by a generalized filled-shell distribution, extrapolated from SWAP observations[38], with cutoff estimated from the upstream SW speed. First, we find that the HTS compression ratio only weakly depends on the PUI density ratio, based on tests assuming ratios of 20 and 30% of the total proton density, for a range of upstream Mach numbers (Figure 3a). Second, the HTS width (foot + ramp) is nearly invariant of the shock parameters (i.e., HTS compression ratio and upstream Mach number; Figure 3b). Therefore, we simplify our analysis by assuming the HTS width is constant. This, however, may not be true in a highly dynamic system with upstream SW turbulence, as suggested by Voyager 2 observations[34]. Thus, we include the shock width as an uncertainty in our analysis by varying it between 1 $L_R$ (Figure 3b) and 1.5 $L_R$[6], where $L_R$ is the upstream advective proton Larmor radius.

**Downstream Particle Distributions**

The PIC simulation results are used to constrain test particle simulations, which we use to simulate the downstream proton distribution for a variety of upstream SW conditions. Realizing the shock compression ratio is minimally affected by the expected range of PUI densities, we ran a set of test particle simulations[6] while varying upstream SW speed, magnetic field, and shock compression ratio (see Methods). The shock width is assumed to be constant in the test particle model, but the width is varied between the width observed by Voyager 2 (3rd crossing of the HTS) and our PIC simulation (Figure 3) and propagated as a form of uncertainty. Examples of the downstream distributions are shown in Figure 4. We focus on the PUI distributions that dominate the ENA fluxes over most of the IBEX-Hi energy range. If we only include PUIs in our test particle model, a rollover of the simulated PUI spectrum appears at ~1 keV, which is not realistic because the spectrum would be dominated by core SW protons at energies below ~1 keV. Because of this artificial rollover, we exclude IBEX ESA 2 data from our analysis (excluding data below the vertical dashed lines in Figure 4), which overlap the rollover and would not provide realistic results because there is no rollover in IBEX observations. Figure 4 shows that the downstream PUI spectral slope changes with shock compression ratio and upstream SW speed, but not significantly with upstream magnetic field. We also note that our test particle simulations assume a quasi-perpendicular shock (angle between the shock normal and upstream magnetic field is ≳75°). The PIC simulation shows that the compression ratio does not significantly change when varying the shock angle values between 75° and 85°. Our MHD simulation shows the HTS is quasi-perpendicular over most of the sky, except near the north and south heliographic poles in just a few pixels where the shock angle is ~60° or smaller (see Figure S1b). In a future study we will consider the quasi-parallel nature of the shock at the poles by varying the shock angle parameter in our test particle code. Because this is just a few pixels near the poles, this will not significantly affect our analysis and results.

**Compression Ratio**

Using the information gathered thus far, we perform a global least-squares and minimization with regularization between the model and data-derived proton fluxes to derive the best-fit HTS compression ratio maps. This involves finding the best compression ratios as a function of location on the HTS surface considering the downstream particle distributions relevant to the IBEX measurement times of interest and using a global MHD simulation to connect each IBEX pixel to multiple HTS foot points. This is necessary to connect IBEX line of sight proton fluxes to modeled proton fluxes downstream of the HTS. A detailed description of this process is provided in Methods.

Figure 5 presents the main results of our work: sky maps and surface plots of the HTS compression ratio, and their uncertainties derived from our analysis. In 2009-2011, solar activity is near the end of solar cycle 23 minimum (accounting for the SW-to-ENA time delay, see Figure S3). Thus, the SW speed is typically faster at high latitudes (Figure 2), yielding higher compression ratios and flatter ENA spectra from the HS. At low-to-mid latitudes, the compression ratio is lower due to the slower SW speeds. The uncertainties are typically <15% for most of the sky except for some small regions, particularly the port side, where the high uncertainties are connected to the regularization uncertainty, which result in different levels of smoothness near the local minimum (in compression ratio) when we vary the SW parameters for our uncertainty propagation analysis. However, IBEX ENA (and thus proton) flux observations show steeper spectra near the flanks, which would imply a lower compression ratio.

Inside the ribbon region, outlined by the white contours, the uncertainties from the minimization process are slightly larger due to the removal of IBEX data (but are typically small). However, because each IBEX pixel is an accumulation of fluxes produced from multiple streamlines connecting to different foot points on the HTS (Figure S6), information surrounding the ribbon gap region is used in the minimization process to fill in the ribbon region.

Interestingly, the port and starboard heliotail lobe regions[60,61], roughly located in longitude ranges -30° to 60° and 90° to 180°, respectively, have the smallest compression ratios. This result reflects the steeper proton spectra observed by IBEX. A possible cause of this behavior, related to the increasing distance to the HTS, is examined in the Discussion section.

Figures 5b, 5d, 5f, and 5h show the HTS compression ratios and their uncertainties for 2014-2016, which approximately reflects SW conditions produced near solar maximum. The SW speeds in this time frame are slower (by a few hundred km s$^{-1}$), particularly at high positive latitudes, and exhibit a noticeable difference in the northern vs. southern hemispheres (Figure 2).

This asymmetry is reflected in the HTS compression ratio primarily because of the relationship between compression ratio and upstream Mach number for quasi-perpendicular shocks[62].

When comparing Figure 5a to 5b (or 5e,g to 5f,h), there is a general decrease in compression ratio, mostly occurring at high northern latitudes, while there is little change below the nose and near the tail direction. The uncertainties of the two maps are, on average, similar, but there are larger uncertainties below the nose and near the direction of Voyager 1 in 2014-2016, and smaller uncertainties at high latitudes in 2014-2016 due to the slower SW speed. The former is due to the higher sensitivity of the compression ratio to the HTS width and interstellar neutral H distribution when the SW speeds were slower, and the latter due to lower uncertainties in SW speed in 2014-2016.

Table 1 shows results in selected directions of the sky. There is no significant change over time in the compression ratio in the nose direction, but there is a statistically significant change near Voyager 1's direction and a possible change in the direction of Voyager 2. The change at Voyager 1 is likely due to the large change in SW speed in the northern hemisphere as the northern polar coronal hole (PCH) was closing[52–54]. The HTS compression ratio near the starboard lobe moderately decreased over time (but still within 1-σ uncertainties), but no significant change is seen from the port lobe. The north and south poles also show notable differences: the compression ratios are similar in 2009-2011, but the compression ratio in the north pole drops significantly due to the PCH closing. Finally, the compression ratio in the tail is quasi-stable over time, though we do note there may be a potential change over time, and the tail compression ratio is typically smaller than the nose compression ratio, especially in 2009-2011.

**Discussion**

By utilizing a sophisticated combination of micro-to-macro modeling tools and global minimization analysis techniques, we have constructed the first all-sky maps of the HTS compression ratio surrounding our solar system. This has been made possible with the IBEX mission's heliospheric ENA imaging capabilities. On average over the sky, the HTS compression ratio typically lies between ~2.55-2.95 in both 2009-2011 and 2014-2016 (±1-σ ranges calculated from the weighted standard deviation of the population), with a mean of ~2.75 over both time periods. These values are slightly smaller than those derived from our global MHD-plasma/kinetic-neutral simulation, but are similar to Voyager 2 observations[35]. Thus, when our results' uncertainties are considered, and that the macro-scale compression ratio observed by Voyager 2 was ~2.5-3.0[63], by including or not the energetic particle precursor[37], we can conclude that our results are reasonably close to Voyager 2. However, the results of our analysis may not strictly be comparable to the timing of Voyager 2's nearly instantaneous, micro-scale crossings of the HTS in 2007 (5 crossings in total), especially considering that the compression ratio changes over time. Also, our methodology does not include energetic particles >10's of keV or their precursor, which, if included, may yield a compression ratio closer to that observed by Voyager 2. Currently, including these effects is beyond our capabilities on a global scale, but should be considered in future studies. Thus, care must be taken when interpreting our results in the context of Voyager observations.

Here we discuss the physical implications of the results, which are illustrated in Figure 6. First, there are clear latitudinal and longitudinal dependencies observed in the HTS compression ratio. The latitudinal dependence primarily comes from the SW speed, which varies between solar minimum and maximum[59]. Faster SW speeds generally yield stronger shocks; thus, the compression ratio is larger at high latitudes in 2009-2011, reflecting solar minimum conditions. The asymmetric evolution of the SW towards maximum activity in solar cycle 24 is visible in Figures 5b and 5f, where the earlier closing of the northern PCH[64] yielded a smaller compression ratio than in the south.

We propose that the longitudinal dependence is driven by (1) a local maximum of interstellar pressure near the nose of the heliopause[1,65,66], (2) the decrease in the SW number density and bulk SW speed towards the flanks where the HTS distance from the Sun is larger than the nose[60,67], (3) slowing of the SW by mass-loading as the distance to the HTS increases along the flanks, and (4) fewer interstellar PUIs in the central tail-ward direction, leading to a smaller plasma pressure and higher upstream magnetosonic Mach number, thus a slightly stronger shock compression[32].

To explain the minimum compression ratios seen at the tailward flanks of the heliosphere, we look at several possibilities beyond what we stated above. First, higher energy particles accelerated along the shock flanks could explain the weaker shock strength along the flanks due to shock mediation. However, using hybrid simulations of the HTS, Giacalone et al.[32] suggested that particles accelerated to energies >10 keV but <1 MeV, which dominate the particle pressure, are likely similar in pressure across the HTS surface, and therefore would not create a spatially-dependent shock compression from the nose to the flanks. Another possibility is that these energetic particles reach maximum energies much higher than 1 MeV along the flanks of the HTS (becoming anomalous cosmic rays; ACRs)[69], but they do not contribute as much pressure as ~10 keV to <1 MeV particles do, and would not be effective at creating an upstream pressure gradient like the one observed by Voyager 2[35,37], and are probably not responsible for the lower compression at the flanks. Finally, it is also possible that ACRs are accelerated in the HS itself, and not at the HTS[70,71], which could also explain Voyager measurements of the "unfolding" ACR spectrum in the middle of the HS, as does the previous theory; while these theories can offer explanations for Voyager observations of ACRs in the HS, the origin of ACR acceleration is not a settled issue.

Therefore, we interpret the compression ratio minimum near the flanks as shown in our results to be a consequence of the slowdown of the SW by mass-loading. This is because the distance to the shock along the flanks is larger than the distance to the HTS near the nose, effectively decreasing the upstream Mach number and weakening the shock. Another revelation from the results can be seen in the differences in the HTS compression ratio between the north and south poles. In 2014-2016, the HTS near the south pole has a significantly larger compression ratio than the north, which may be caused by faster SW in the south due to the larger PCH, and the disappearance of the PCH in the north, as shown by Figure 2 for the second time period.

It is important to note that while we include particle acceleration in the HS in our analysis via stochastic acceleration from the Parker transport equation[9], as well as adiabatic heating via compression of the plasma along flow streamlines, there is uncertainty to the velocity diffusion coefficient's spectral index. Therefore, we propagate the uncertainty of the spectral index in our compression ratio results. As we describe in more detail in Methods, there is no consensus on what mechanism is primarily operating in the HS and a recent study showed additional heating in the HS may not be needed[7] to explain the "gap" between modeled and observed ENA fluxes identified in recent years[10,67,72,73]. However, it is also believed that particles are likely heated/accelerated by Alfvénic or compressible turbulence in the HS, considering Voyager observations of turbulence in the HS[74–76], which likely originated from the transmission of SW turbulence and/or current sheets across the HTS into the HS[77–79,75,80]; thus, we included a potential HS heating mechanism in our methodology based on successful modeling fits to IBEX data[9,81]. By using our method of velocity diffusion in the HS based on previous studies that fit to IBEX-Lo and IBEX-Hi observations of ENA emissions from the central heliotail, the new simulated ENA maps match better to IBEX observations (Figure S8) compared to some previous studies[10,67,82], at least in the forward hemisphere of the heliosphere (except longitudes ~30° to ~120°, which is centered on the heliotail).

Finally, the upcoming Interstellar Mapping and Acceleration Probe (IMAP) mission[83], which is expected to launch in 2025, will provide ENA measurements with better statistics, over a broader range of energies, and improved temporal cadence. IMAP ENA data will allow us to improve our determination of the HTS compression ratio as a function of space and time, and better understand the role of energetic particle mediation on the tail-ward flanks of the shock with the comprehensive suite of in situ particle instruments onboard the spacecraft. This manuscript provides a necessary advancement for future studies to improve upon our understanding of the HTS, particularly with IMAP measurements.

## Methods

**Propagating SW/PUI Plasma to the HTS**

We propagate the SW $H^+$, $He^{++}$ and interstellar $H^+$ and $He^+$ PUI distributions to the HTS using multi-species equations[15]. To summarize here, we solve the following equations coupled by photoionization and charge exchange source terms:

$$\frac{1}{r^2}\frac{d}{dr}(r^2\rho u) = \sum_{i=1}^{4} S_i^\rho, \qquad (1)$$

$$\frac{1}{r^2}\frac{d}{dr}(r^2\rho_i u) = S_i^\rho, \qquad (2)$$

$$\frac{1}{r^2}\frac{d}{dr}(r^2\rho u^2) = S^m - \frac{dp}{dr} - \frac{B_\varphi^2}{\mu_0 r} - \frac{B_\varphi}{\mu_0}\frac{dB_\varphi}{dr}, \qquad (3)$$

$$u\frac{dp_i}{dr} = S_i^p - \gamma p_i \frac{1}{r^2}\frac{d}{dr}(r^2 u), \qquad (4)$$

$$\frac{1}{r^2}\frac{d}{dr}(r^2 B_r) = 0, \qquad (5)$$

$$B_\theta = 0,$$

$$\frac{1}{r}\frac{d}{dr}(rB_\varphi u) = 0,$$

where these equations solve for (1) total mass flux, (2) individual mass fluxes of SW $H^+$, $He^{++}$ and interstellar $H^+$ and $He^+$ PUI separately, (3) total momentum flux, (4) individual internal pressures for each ion species, and (5) interplanetary magnetic field with radial ($B_r$) and tangential ($B_\varphi$) components whose total magnitude at 1 au is extracted from OMNI data and initialized at 1 au using the Parker Spiral equations[84]. The angle between the radial direction and the magnetic field direction at 1 au is assumed to be 45°, a typical average angle consistent with OMNI data. The source terms ($S^\rho, S^m, S^p$) on the right side of Equations (1)-(4) can be found in the Methods section of our previous study[15]. The source terms include the effects of photoionization and charge exchange, using a 3D spatial distribution for neutral H derived from our global MHD model, including the recent update to neutral H density from SWAP[85,86], and for interstellar He derived from the "cold" model[87,88].

The global MHD model simulates the SW-LISM interaction by solving the single fluid MHD equations that are then coupled, through source terms, to neutral H atoms which are solved using Boltzmann's equation. The inner SW boundary conditions (defined at 1 au and extrapolated to the inner boundary of the simulation at 10 au) of the global MHD model are extracted from the OMNI database and Ulysses observations, averaged over 2004-2009, which is chosen based on the expected average return time for ENAs measured in 2009-2016[51]. In the low latitude, slow SW, the inner boundary conditions are extracted from the OMNI database: speed = 449 km s$^{-1}$, density = 6.53 cm$^{-3}$ (for both protons and alphas as a single fluid), temperature = 1.05×10$^5$ K, and radial magnetic field = 37.4 µG (assuming a Parker spiral). In the high latitude, fast SW, the inner boundary conditions are extracted from Ulysses data taken during its third fast polar scan in 2007: speed = 743 km s$^{-1}$, density = 2.23 cm$^{-3}$ (both protons and alphas), temperature = 2.98×10$^5$ K, and radial magnetic field = 34.7 µG. The latitude of separation between the slow and fast SW is |±37°|, based on Ulysses observations[59]. The inner boundary conditions are then extrapolated to 10 au assuming a Parker spiral for magnetic field and adiabatic expansion of the plasma from 1 to 10 au from the Sun.

The outer boundary conditions of the MHD model, set at 1000 au from the Sun, are extracted from IBEX and New Horizons' SWAP observations: speed = 25.4 km s$^{-1}$ [89], total effective plasma density = 0.09 cm$^{-3}$ (interstellar H$^+$ density is ~60% or 0.054 cm$^{-3}$, and the remainder is interstellar He$^+$ with 10% of the density, 0.009 cm$^{-3}$ [90], but 40% of the dynamic pressure – note that interstellar neutral He$^+$ is not solved as a separate fluid, but rather that in the MHD/Boltzmann charge exchange source terms the plasma is considered to be comprised of this H$^+$ and He$^+$ mixture, thus approximating the presence of interstellar He$^+$ through charge exchange), neutral H density = 0.17 cm$^{-3}$, temperature = 7500 K[89], and magnetic field strength 2.93 μG and orientation (227.28°, 34.62°)[91]. The neutral H density is chosen such that the filtration of interstellar neutral H through the front of the heliosphere yields H densities consistent with New Horizons' SWAP observations[85].

We propagate the SW/PUI parameters from 1 au to the HTS, where the HTS location is assumed to be that from the same global MHD simulation used to produce the neutral H density distribution[15] (see Figure S1a), but we scale up the distances by ~10% to match the averaged observed distances in the Voyager 1 and 2 directions. Because the HTS location is variable over time, and its distance is uncertain, we propagate a 1-σ uncertainty of 5% to the distance to the HTS in our analysis (see "Finding the Best-fit Compression Ratio at the HTS"). Once the SW/PUI parameters upstream of the HTS are found, we use these to run a range of test particle simulations with different upstream parameters encompassing the results from solving Eq. (1)-(5). The details of the test particle simulations are described in "Deriving Model Proton Fluxes Downstream of the HTS".

Moreover, we note that our MHD simulation overestimates the thickness of the HS by ~15 au in the Voyager 1 and 2 directions as compared to the Voyagers' observations. However, the distances to the middle of the HS in the Voyager 1 and 2 directions are, on average, consistent with Voyager 1 and 2 observations (i.e., the HTS distance is smaller than observed, and the HP distance is larger than observed). Thus, in terms of current capabilities, this is a reasonable simulation to use for our analysis, particularly for the flow streamlines and SW-ENA time delays.

For simplicity in our analysis, we create two sky maps of the HTS compression ratio, one corresponding to IBEX data taken in years 2009 through 2011, and another for 2014 through 2016. These time periods approximately correspond to solar minimum and maximum conditions in solar cycle 24, respectively (accounting for the time delay for changes to occur in ENAs observed at 1 au), and the combination of 3 annual maps per period improves the statistics of the analysis. The time delay between SW outflow measured at 1 au and IBEX measurements at 1 au can range anywhere between ~1.5 and >7 yr, depending on the ENA energy and direction in the sky. We describe ways of handling this complexity in "Estimating Time Delays between SW and ENA Measurements".

We solve Eq. (1)-(5) starting at 1 au using SW conditions for each year from 1998.5 to 2016.5 (the IPS SW speed model has one solution per year[52]) to cover the possible time delays between SW and ENA detection yielded by the simulation. This yields SW/PUI plasma properties upstream of the HTS for a large range of years after propagation to the HTS. Example plots of SW speed, density, and magnetic field upstream of the HTS are shown in Figure S2.

**Estimating Time Delays between SW and ENA Measurements**

A better way than assuming a single delay time for all ENA energies and pixels in the sky is to first use a global MHD simulation to estimate the total time delay per pixel and ESA passband of IBEX. Therefore, we utilize a global MHD simulation[15], which includes the higher neutral H density found from SWAP observations[85] and adapts the interstellar plasma density to include the presence of interstellar He$^+$ in the charge exchange source terms. Moreover, this simulation also constrains the interstellar densities to require that the middle of the simulated HS averaged in the Voyager 1 and 2 directions is the same as the middle of the HS averaged over Voyager 1 and 2 observations. The interstellar magnetic field at the outer boundary of the simulation is the same found by fitting to the IBEX ribbon[91]. Using this simulation, for each pixel in the sky, we integrate backwards the time it takes an ENA at a specific energy within an IBEX ESA passband to go back to its origin in the HS (first starting immediately downstream of the HTS), for the HS plasma at that point to flow back to the HTS, and for the supersonic SW to propagate from that position back to 1 au. Note that because the observed ENA signal is a line-of-sight integration, we calculate multiple delay times by assuming the ENA may come from anywhere in the HS along the line of sight, tracing back to the HTS and Sun at multiple foot points, yielding multiple delay times. We then find the mean delay time by averaging these delay times weighted by the local proton flux in the HS, i.e., higher proton fluxes producing higher ENA emissions yields a larger weight. This method of weighting is the same as that used by Zirnstein et al.[51], see specifically their Eq. (10)-(11). As shown by Zirnstein et al.[51], the local proton flux, $f_p v^2/m_p$, is derived as a local weight from the line of sight integrated ENA flux equation in order to calculate the ENA and proton flux in the HS plasma frame per pixel in the sky (see Equations (4)-(11) in Section 2.2 of their paper for more details). We then repeat this procedure over a range of ENA energies covering the IBEX ESA passband, and weight-average them using the ESA response function[92]. This is repeated for each ESA of IBEX-Hi (except ESA 2). Example sky maps of the total delay times for ESAs 3 and 6 are shown in Figure S3.

**Deriving Model Proton Fluxes Downstream of the HTS**

With a set of plasma conditions upstream of the HTS for a range of years ("Propagating SW/PUI Plasma to the HTS"), and the total time delay between SW and ENA measurements at 1 au ("Estimating Time Delays between SW and ENA Measurements"), we subtract the total time delay in the IBEX spacecraft ram frame for each year in the two time periods and find the upstream SW properties corresponding to the ENA measurements. We then average the plasma properties over time that correspond to the two IBEX time periods. Examples of the time-averaged upstream plasma conditions are shown in Figures 2 and S4. One can see the effects of the spacecraft's Sun-pointed spinning and the abrupt change in plasma conditions at longitude 180° because of the disjoint in time while only using ram frame observations.

Using this information, we know that the ranges of upstream magnetic field magnitude and SW speed are [0.001, 0.07] nT and [200, 600] km s$^{-1}$, respectively. We do not need to know the upstream plasma density, because our analysis only uses normalized downstream fluxes. We also assume that the PUI-to-SW proton number density ratio is 0.25 everywhere, based on their small effect on the shock structure and compression ratio as seen in PIC simulation results[31], shown in Figure 3. As one can see, the compression ratio of the simulated HTS only weakly depends on the PUI density ratio and are small compared to the uncertainties of our results, and thus for simplicity we ignore its effects on our analysis. The PUI-to-SW proton number density ratio does change moderately across the sky. Estimates from, e.g., Zirnstein et al.[67], show that the ratio can change between ~0.2 and 0.3 (see their Figure 2), with the lowest ratio near the nose of the heliosphere and the highest between the poles/flanks and tail of the heliosphere. We will relax this assumption of a constant density ratio in a future study.

To derive model proton fluxes downstream of the HTS, we use a test particle simulation[6] over the range of parameters found above, where we assume the upstream velocity distribution is a filled shell with adiabatic cooling index set to 2.9, extrapolated from New Horizons' SWAP observations halfway to the HTS[38]. Specifically, we run the test particle simulation for all combinations of three variables: (1) upstream magnetic field magnitude (assuming a perpendicular shock) with values [0.001, 0.0355, 0.07] nT; (2) upstream SW speed with values [200, 300, 400, 500, 600] km s$^{-1}$; and (3) HTS compression ratio with values [2.0, 2.5, 3.0, 3.5, 4.0]. Examples of the downstream distribution results are shown in Figure 4. We do not run test particle simulations for different SW or PUI densities because these just scale the downstream distribution by constant factors, and our derivation of the optimal compression ratio involves minimizing chi-square between normalized IBEX and model proton fluxes ("Finding the Best-fit Compression Ratio at the HTS"). The downstream distribution is linearly interpolated between the values shown above for different upstream SW conditions.

Figure 4 shows how the downstream PUI distribution function depends on the upstream magnetic field, SW speed, and HTS compression ratio. For example, Figure 4a demonstrates how the distribution becomes flatter at energies above ~1 keV and higher in intensity. As the upstream SW speed increases for a single compression ratio (Figures 4a-c), the distribution shifts towards higher energies because the downstream flow is moving faster away from a solar inertial observer (or, e.g., IBEX). Moreover, as the upstream SW speed increases, the compression ratio increases and acceleration from the cross shock electric field is stronger. Also notice that in each panel we plot several solid and dashed vertical lines. The solid vertical lines show the central energy of ESA 2 (~0.7 keV) minus ESA 2's half-width at half-maximum (HWHM) for each downstream distribution curve. The dashed vertical lines are similar, except they correspond to ESA 3 minus it's HWHM. We show these lower 'edges' of the ESA passbands because they are close to the rollover in the PUI distribution. We do not include core SW protons in our test particle simulation because they are not accurately transported across the HTS in our simulation due to their sensitivity to the micro-structure (at scales much smaller than for PUIs) of the simulated shock. Their exclusion from our simulation creates an artificial rollover at and below ~1 keV and thus cannot reproduce IBEX ESA 2 observations, which do not have this rollover. Therefore, we exclude ESA 2 data from our analysis, whose energies overlap where the rollover occurs.

Figures 4d-f show how the downstream PUI distribution depends on the upstream magnetic field, for different HTS compression ratios in each panel. Clearly, there is no visible difference in the results as a function of upstream magnetic field. This is because for a single HTS compression ratio and upstream SW speed, the gain in energy per particle depends on the ratio of the mean turbulent magnetic field to the mean field power, i.e., $(\delta B/B)^2$, which is a constant in our model.

Finally, we compare our test particle results to Voyager 2's Low Energy Charged Particle (LECP) observations, as shown in Figure S5. We set the compression ratio to 2.5 (approximately the large-scale HTS compression ratio during Voyager 2's crossing), and show results for upstream flow speeds of 300 km s$^{-1}$ and 400 km s$^{-1}$, which is the range of speeds that Voyager 2 within ~0.7 au of the HTS. Despite the lack of statistics at energies above ~15 keV in our test particle model, our results are consistent with the observations.

**Mapping Positions on the HTS to each IBEX Pixel**

ENAs created along a particular IBEX line of sight do not originate from a single point on the HTS, but rather from multiple points depending on the bulk flow pattern in the HS[49]. In general, most ENAs integrated over a line of sight originate from PUIs that crossed the HTS within approximately 10°-30° of the line of sight, but this also depends on the region of the sky and deflection of the HS flow downstream of the HTS. Therefore, in the derivation of the best-fit compression ratio for a particular IBEX line of sight (or pixel) in the sky (see "Finding the Best-fit Compression Ratio at the HTS"), we minimize over the normalized sum of multiple HTS foot points that contribute to the pixel. We utilize HS plasma flow streamlines from our global MHD simulation[15] to mimic this effect. An example sky map showing several pixels in the sky and where they connect to the HTS is shown in Figure S6. The weighting given to each HTS position (colored pixels) is based on the contribution it makes to the total ENA flux from the IBEX pixel (black diamond symbols), i.e., the ENA flux produced from line of sight (LOS) element $dl$ by the local HS proton distribution which propagated from a point on the HTS surface. The reversal of this, i.e., the contribution of one HTS foot point to multiple IBEX pixels, is used in our global minimization scheme.

**Effects of Velocity Diffusion, Adiabatic Heating, and Charge Exchange in the HS**

Because the IBEX proton fluxes represent the line of sight-averaged flux in the HS for each IBEX ESA, we must "undo" or deconvolve the effects of propagation through the HS to properly compare the observed proton fluxes to the modeled proton fluxes just downstream of the HTS. We do this by solving the Parker transport equation with charge exchange source terms on the large scale plasma flows obtained from our global heliosphere model, similar to our previous work[9,81], but this time in all directions of the sky. Each line of sight has ENA emissions produced at different distances through the HS, with each having a plasma flow streamline connected back to a different point on the HTS. Therefore, for each of these streamlines, first we trace the flow back to the HTS via a 2nd order Runge-Kutta (midpoint) method. Upon reaching the HTS, we initialize a downstream proton distribution that is, on average, close to what is expected downstream of the HTS based on prior studies (i.e., a kappa distribution[7,51] with kappa index = 2.2; the kappa index is varied between 2 and 2.4 to estimate the 1-σ uncertainty of the index in the final results). While it would make more sense to have a spatially varying kappa index downstream of the HTS, currently it is not possible to determine what they should be immediately downstream of the shock. Therefore, instead, we take a nominal value for all locations on the HTS and test different kappa indices that are then propagated as uncertainties. We note that using kappa indices of 2 and 2.4 do not change the compression ratio results significantly. Then, we solve the Parker transport equation with source terms, forward in time to the ENA emission position, as shown below in finite difference form (see Zirnstein et al.[9] for more details):

$$f_j^{n+1} = f_j^n + \frac{\Delta s e^{-3w_j}}{u_p \Delta w^2} \left[ D_{j+\frac{1}{2}} e^{w_{j+\frac{1}{2}}} (f_{j+1}^n - f_j^n) - D_{j-\frac{1}{2}} e^{w_{j-\frac{1}{2}}} (f_j^n - f_{j-1}^n) \right]$$
$$+ \frac{\Delta s}{3 u_p} (\nabla \cdot \boldsymbol{u}) \left[ \frac{f_{j+1}^n - f_{j-1}^n}{2 \Delta w} \right] + \frac{\Delta s}{u_p} \left[ \eta_j^n f_{H,j}^n - \beta_j^n f_j^n \right], \tag{6}$$

where $w = \ln(v)$ and $\Delta s$ is the step size along the flow streamline. The second term on the right-hand side is represents velocity diffusion, with diffusion coefficient $D(v) = D_0 v^\alpha$ (where $D_0$ is the diffusion amplitude and $\alpha$ is the spectral index), the next is flow divergence (causing adiabatic heating/cooling) where $(\nabla \cdot \boldsymbol{u})$ is calculated using the global MHD simulation's change in density compression along flow streamlines, and the last is the charge exchange source term. This gives us a final proton distribution/flux at the ENA emission point along the line of sight in question. Note that (1) the typo in Equation (11) in Zirnstein et al.[9] with the missing $\Delta s$ in the charge exchange source term is fixed here, and (2) we have simplified the production of protons by separating out the neutral H distribution $f_H$ and the charge exchange production rate $\eta$ in Equation (2) of Zirnstein et al.[9]. This does not significantly affect our results because the production of protons in the HS are injected at speeds in the plasma frame at energies around ~100 eV. Equation (6) is solved using a "forward-time, central-difference" method. Therefore, the step size along the streamline, $\Delta s$, must be sufficiently small to maintain stability. We have found that a step size of ≤0.02 au is sufficient. The range of speeds over which we solve Equation (6) is 1 to 6200 km s$^{-1}$, or ~0.005 eV to 200 keV, in natural log-space, with 150 bins.

We calculate the ratio of the final distribution function over the initial distribution just downstream of the shock at a specific particle speed $v$, i.e., $F(v) = f_f(v)/f_i(v)$, for each streamline (where $v$ is the desired ENA speed to measure and the initial distribution $f_i(v)$ is described by a kappa = 2.2 distribution). Each streamline ratio is averaged along the line of sight to find the "best" average change in the distribution. Next, we divide the IBEX proton flux along the line of sight by this ratio (see Equations 12 and 13), effectively "undoing" the effects of velocity diffusion, adiabatic heating, and charge exchange as the distribution evolves through the HS. This is possible because the flux is proportional to the distribution. Examples of this ratio across the sky are shown in Figure S7. We note that taking an average of this ratio along the line of sight may introduce unknown systematic uncertainties; however, our current methods are the best that we can provide at this time.

The diffusion coefficient, $D(v) = D_0 v^\alpha$, is based on a previous study[9] where the IBEX-Lo and IBEX-Hi spectra were fit by chi-square minimization to find the best values for $D_0$ and $\alpha$. By fitting a parabola to the minimum of each spectral index case in the left panel of Figure 4 from Zirnstein et al.[9], we find the best fit values with minimum chi-square are $D_0 = 8.18 \times 10^{-9}$ km$^2$ s$^{-3}$, and $\alpha = 1.31 \pm 0.20$, in the central tail direction. The uncertainties of the diffusion amplitude $D_0$ and the spectral index $\alpha$ are not independent but reflect their tight linear correlation in log space (see the narrow parabolas in the left panel of Figure 4 from Zirnstein et al.[9]). Therefore, they need to be considered jointly as one uncertainty. We parameterized this uncertainty by the $\alpha$ uncertainty, and for the 1-σ uncertainty of $\alpha$ we use the corresponding $D_0$ value based on the correlation between the parameters. See "Finding the Best-fit Compression Ratio at the HTS" for information on the propagated uncertainties. We note that while this diffusion coefficient was derived from a single direction in the sky, the amplitude of $D_0$ changes with direction in the sky, as described below. The spectral index $\alpha$ = 1.31 is, interestingly, halfway between the diffusion index for incompressible/Alfvén turbulence and compressible/wave-like turbulence[77], both of which are observed in the HS[76].

Because our diffusion coefficient was only derived in the central tail direction, we make a few scaling modifications to apply it to other directions in the sky. First, we set the nominal $\alpha$ = 1.31 across the sky, because we include its uncertainty in our analysis. Second, we set $D_0 = 8.18 \times 10^{-9}$ km$^2$ s$^{-3}$ as the nominal value in the central tail pixel but scaled by a certain factor. This scaling factor accounts for the fact that the heliosphere simulation used to derive the diffusion coefficient in our prior study has different SW/VLISM properties, yielding a different distance to the HTS in the central tail direction. One might set the scaling factor to the inverse square of the distance to the shock[93] (which is approximately proportional to the ratio of turbulence to mean field power, $(\delta B/B)^2$) with respect to that in the central tail direction. This means that directions where the HTS is closer to the Sun would have a higher level of turbulence because of less time spent to reach the shock (and less time for the dissipation of turbulence). The scaling factor in this case would be $(r_{1au}/r(\Omega))^\omega$ where $\Omega$ signifies any particular direction in the sky and $\omega = 9(1 + \Gamma)/4$ is the power law spectral index (see Equation (35) in Zank et al.[93]). However, this formulation only works for a single inner boundary SW speed. Because we have different speeds as a function of longitude (because of PUI mass-loading) and latitude, we update this ratio to be a function of SW propagation time to the HTS rather than distance, yielding $(\tau_{tail}/\tau(\Omega))^\omega$. In this way we have replaced $r_{1au}$ with the time for travel to the HTS in the central tail direction, $\tau_{tail}$. Thus, $D_0$ in the central tail pixel direction is scaled by a factor $\sim(\tau_{tail,ref}/\tau_{tail})^\omega$, where $\tau_{tail,ref} = 1.34$ yr is the reference mean time from Zirnstein et al.[9] where the diffusion coefficient was originally derived, and $\tau_{tail} = 1.19$ yr is the mean time in the heliosphere simulation used in this manuscript. The scaling index $\omega \cong$ 2-3, where $\omega$ is closer to 3 within and near the ionization cavity, and $\omega$ is closer to 2 farther beyond the ionization cavity[93]. By estimating the mean spectral index of the evolution of Voyager 1, Voyager 2, and Pioneer 11 observations of $(\delta B/B)^2$, shown in Figure 4 in Zank et al.[93], we find $\omega \cong 2.5$ is a good approximation from ~20-40 au, making the first scaling factor $(\tau_{tail,ref}/\tau_{tail})^{2.5}$. These results are also quite similar to the results obtained from a more recent, sophisticated model of turbulence transport[94].

We then must apply a second scaling factor for $D_0$. This factor is based on the time it takes the SW to reach the HTS in directions of the sky other than the central tail direction in the current simulation (because the distance to the HTS and SW speed is not the same in all directions). Thus the second scaling factor is $(\tau_{tail}/\tau(\Omega))^{2.5}$, where $\tau(\Omega) = \langle r(\Omega)/u_p(\Omega) \rangle$ is the mean time it takes the SW to reach the HTS in a certain direction of the sky, and $\tau_{tail}$ is the mean time in the central tail direction. Thus, the faster the SW speed ($u_p(\Omega)$) and/or closer distance to the HTS ($r(\Omega)$), the larger this ratio is, and the larger the turbulence power.

Finally, the total scaled diffusion amplitude, $D_0'(\Omega)$, can be written as

$$D_0'(\Omega) = D_0 \left(\frac{\tau_{tail,ref}}{\tau_{tail}}\right)^{2.5} \times \left(\frac{\tau_{tail}}{\tau(\Omega)}\right)^{2.5} = D_0 \left(\tau_{tail,ref}/\tau(\Omega)\right)^{2.5}. \tag{7}$$

where $D_0'(\Omega)$ replaces $D_0$ in Equation (6).

We note that most global heliosphere models of IBEX ENA measurements underestimate the intensities by a factor of ~2-3[67,82,10], either globally or in certain regions of the sky. Because these models usually only include adiabatic heating effects, it suggests that further particle heating may be occurring in the HS which could explain this discrepancy. Some studies have suggested particle heating/acceleration by reconnection[70,95,96], turbulence[71,77], or shocks[97] may also occur in the HS. Currently there is no consensus which mechanism, or mechanisms, may dominate over others, and how to implement in a global heliosphere model. Heating by shocks passing through the HS, however, would not affect our analysis because this primarily raises the intensity of ENAs in the IBEX-Hi energy range, but does not significantly change the spectral slope[97]. Other mechanisms that primarily accelerate particles producing suprathermal tails is beyond the energy range considered here. Finally, a recent study[7] demonstrated that particle acceleration at the HTS based on test particle simulations may be sufficient to explain the IBEX-Hi ENA spectrum without the need for additional, non-adiabatic heating in the HS. However, that determination was made only for IBEX ENA data in the Voyager 2 direction, and therefore is limited in its conclusions regarding global heating in the HS.

In our study, we include a velocity diffusion process of particles as they travel along the HS plasma flow streamlines, with a velocity diffusion coefficient based on previous fits to IBEX data[9]. The results of the prior study suggested that the spectral index of the diffusion coefficient lies between acceleration by Alfvénic turbulence ($\alpha = 2/3$) and compressible turbulence ($\alpha = 2$)[77], i.e., close to the value of 1.3 that we employ through the HS, as noted before. However, the spectral index has an uncertainty (see "Finding the Best-fit Compression Ratio at the HTS"), which is propagated through our results. As shown in Figure S8, the inclusion of velocity diffusion greatly improves the comparison with IBEX GDF-separated data. The old simulation results (Figure S8a) greatly underestimate the observations (Figure S8c-e) by more than a factor of 2. By including velocity diffusion in the HS, the new simulation results (Figure S8b) are increased by approximately a factor of 2 or more, lying within the uncertainty range of the observations. We note that the high flux from heliotail at mid to high latitudes (except the central downwind direction) go beyond 100 flux units. This is primarily due to the steady-state nature of the MHD simulation used to create these maps, where consistent fast SW flows down the tail. In reality, it would be a mixture of slow to fast SW[67].

**Compton-Getting Correction for Realistic HS Flow Frame**

It is important to note that the IBEX proton fluxes are in the simulated, steady-state HS plasma frame[51]. The upstream SW conditions and downstream flow speeds may, therefore, be different than what we expect based on ACE/Wind and IPS observations. Moreover, when performing our analysis, we must vary the shock compression ratio for all combinations of upstream SW speeds – thus, the IBEX proton fluxes are not likely to be in the correct reference frame at each iteration of the chi-square minimization routine. Therefore, we calculate Compton-Getting correction factors for the IBEX proton fluxes for each direction in the sky, ESA, time period, and compression ratio tested in our analysis. These correction factors are multiplied to the IBEX proton fluxes and energies in the minimization routine for the appropriate variable value. The correction factor is calculated as

$$c(\Omega, E_p, t, r_{HTS}) = 1 - 2\frac{\Delta u}{v_p}\cos[\langle\beta(\Omega, E_p)\rangle] + \left(\frac{\Delta u}{v_p}\right)^2, \quad (8)$$

where $c$ is the correction factor as a function of direction in the sky ($\Omega$), central energy of the ESA in the simulated plasma frame ($E_p$), time period $t$, and shock compression ratio $r_{HTS}$. The particle speed in the simulated HS plasma flow frame is $v_p = \sqrt{2E_p/m_H}$. The variable $\Delta u$ is the speed difference between the MHD simulation's HS plasma frame and the desired frame based on the SW speed upstream of the HTS and the HTS shock compression ratio. We estimate this as

$$\Delta u(\Omega, E_p, t, r_{HTS}) = \frac{u_{SW,u}(\Omega, E_p, t)}{r_{HTS}} - \langle u_{MHD,d}(\Omega, E_p)\rangle, \quad (9)$$

where $u_{SW,u}(\Omega, E_p, t)$ is the SW speed upstream of the HTS, and $\langle u_{MHD,d}(\Omega, E_p)\rangle$ is the SW speed immediately downstream of the HTS from the MHD simulation. We note that the downstream speed from the MHD simulation is averaged over all streamlines that connect back to the HTS from different radial increments along the IBEX line of sight direction, $\Omega$, and weight averaged along the line of sight, using similar averaging methodology as Zirnstein et al [51]. Equation (9) simplifies the problem by assuming that the differences in flow speed do not influence the plasma flow patterns in the HS. This effectively assumes that the two different plasma frames are parallel (when $\Delta u > 0$) or anti-parallel (when $\Delta u < 0$).

Because the two reference frames for the transformation are parallel (or anti-parallel), the angle $\beta$ can be calculated as the angle between the proton velocity in the MHD-simulated HS plasma frame, $v_p$, directed towards IBEX, and the simulated HS plasma frame velocity, $u_{MHD}$, such that

$$\beta(r, \Omega, E_p) = \cos^{-1}\left[\frac{v_p \cdot u_{MHD}}{|v_p||u_{MHD}|}\right],$$

$$\langle \beta(\Omega, E_p) \rangle = \frac{\int_{r_{HTS}}^{r_{HP}} \beta(r, \Omega, v_p) \omega(r, \Omega, v_p) dr}{\int_{r_{HTS}}^{r_{HP}} \omega(r, \Omega, v_p) dr}. \quad (10)$$

Angle $\beta$ is calculated at every position along each IBEX line of sight and then weight averaged to produce $\langle \beta \rangle$.

Figures S9 and S10 show examples of the Compton-Getting correction factor for the all-sky maps at ESA 3 and 6 and $r_{HTS} = 2.5$ and 3.5. The most distinct feature in all panels is that the correction factor is <1 at high latitudes, especially for larger $r_{HTS}$ and for the second time period (2014-2016). The reason for this is that the (steady-state) MHD simulation assumed fast SW speeds at 1 au of 743 km s$^{-1}$ at latitudes >|±37°|, whereas the speeds derived from IPS-derived models[52] are slightly lower (note that IPS speeds are shifted down at all latitudes to match OMNI). This effectively means the plasma frame is moving too fast away from the observer in the simulation, and therefore IBEX-derived proton fluxes and corresponding energies need to be decreased to compensate. The reason why this is more pronounced in 2014-2016 is because the solar cycle is approaching solar maximum in the time-delayed ENA source frame, thus the average SW speeds at high latitudes are much smaller than the steady-state, solar minimum-like simulation.

The second distinct feature in these maps is that the correction factor is sometimes above 1 at lower latitudes, particularly near the nose and tail. It becomes less than 1 for large $r_{HTS}$, similar to the reasons stated above. The correction factors above 1 at low latitudes suggest either the MHD-simulated SW speeds were too low and/or the MHD-simulated compression ratio was too high (compared to the value assumed in the calculation of each specific map). Overall, by performing these corrections we aim to partially remove the influence of the MHD simulation's assumptions for SW speed and HTS compression ratio on our results. We tested the sensitivity of our compression ratio results when using these correction factors vs. not using them and found there is no significant change compared to the total uncertainties of our results.

**Finding the Best-fit Compression Ratio at the HTS**

To find the best-fit HTS compression ratios as a function of direction in the sky, we first start with the proton distribution in the HS averaged over the line of sights derived from IBEX observations. Let $j_{t,p,s} \pm \delta j_{t,p,s}$ denote the derived flux and its uncertainty for year $t$, pixel $p$, and ESA step $s$. While the fluxes are derived for each observed ESA step, the energy in the plasma frame $E_{t,p,s}$ differs from the nominal central energy in the heliocentric frame due to the Compton-Getting effect[98–100]. For simplicity, we enumerate pixels using a single integer $p = 1, 2, \ldots, 1800$. In this study, we accumulate the fluxes over several years corresponding to time period $\tau$ (i.e., 2009-2011 and 2014-2016). In general, we calculate the energy and the average flux over this period as well as its uncertainty from the following formulae:

$$\bar{E}_{\tau,p,s} = \frac{1}{|\tau|} \sum_{t \in \tau} E_{t,p,s}, \quad (11)$$

$$\bar{J}_{\tau,p,s} = \frac{1}{|\tau|} \sum_{t \in \tau} j_{t,p,s} / F_{t,p,s}, \quad (12)$$

$$\delta \bar{J}_{\tau,p,s} = \frac{1}{|\tau|} \left( \sum_{t \in \tau} \delta j_{t,p,s}^2 / F_{t,p,s}^2 \right)^{\frac{1}{2}}, \tag{13}$$

where $|\tau|$ denotes the number of years included in the sum. We omit missing data points in these sums. As mentioned earlier, the IBEX fluxes and their uncertainties are divided by another scaling factor, the distribution ratio that includes the effects of velocity diffusion, adiabatic heating, and charge exchange, which we define here as $F_{t,p,s}$.

The proton spectrum observed along each line of sight includes contributions from multiple streamline foot points at the HTS. The contribution along line of sight $p$ from point $q$ on the HTS in ESA step $s$ is denoted as $m_{s,p,q}$. Note that the lines of sight are numbered with index $p$, while points at the HTS are numbered with $q$. Because we are only interested in relative contributions, we normalize the weights using the following formula:

$$\widetilde{m}_{s,p,q} = \frac{m_{s,p,q}}{\sum_{q=1}^{1800} m_{s,p,q}}. \tag{14}$$

This formula ensures that the weights for a given line of sight sum up to 1.

In our analysis, we compare the derived proton fluxes from IBEX with modeled fluxes as a function of compression ratio. Let $g_{\tau,q,s}(E,R)$ denote the modeled proton flux at point $q$ at the HTS, in energy step $s$, for conditions corresponding to period $\tau$ at energy $E$ and compression ratio $R$. As stated earlier, a relative Compton-Getting factor $c_{\tau,p,s}(R)$ needs to be applied to the derived flux and corresponding energy in each line of sight (for details, see "Compton-Getting Correction for Realistic HS Flow Frame"). These correction factors are multiplied to the average flux and energy in Equations (15) and (16).

Based on the above, we want to minimize the following least-squares expression:

$$\chi_{\text{LS},\tau}^2(\boldsymbol{R}, \boldsymbol{a}) = \sum_s \sum_p \frac{\left( c_{\tau,p,s}(R_p') \bar{J}_{\tau,p,s} - a_p \sum_q \widetilde{m}_{s,p,q} g_{\tau,q,s}(c_{\tau,p,s}(R_p') \bar{E}_{\tau,p,s} R_q) \right)^2}{\left( c_{\tau,p,s}(R_p') \delta \bar{J}_{\tau,p,s} \right)^2}. \tag{15}$$

It can be rewritten in an equivalent form:

$$\chi_{\text{LS},\tau}^2(\boldsymbol{R}, \boldsymbol{a}) = \sum_s \sum_p \frac{\left( \bar{J}_{\tau,p,s} - c_{\tau,p,s}^{-1}(R_p') a_p \sum_q \widetilde{m}_{s,p,q} g_{\tau,q,s}(c_{\tau,p,s}(R_p') \bar{E}_{\tau,p,s} R_q) \right)^2}{\left( \delta \bar{J}_{\tau,p,s} \right)^2}. \tag{16}$$

In the above expression, we seek the best-fit compression ratio and normalization factor vectors: $\boldsymbol{R} = \{R_q\}_{q=1,\dots,1800}$, $\boldsymbol{a} = \{a_p\}_{p=1,\dots,1800}$. Note that the IBEX proton flux maps exclude pixels affected by the IBEX ribbon flux (see Figure 1 in the main text). Therefore, we remove them from the sum in Equation (16). Consequently, the normalization factors for these pixels $a_p$ are not constrained in our study and remain undefined. They are, however, not the primary interest of this study, and thus we do not attempt to estimate these values. The Compton-Getting correction in Equation (16) depends on the effective compression ratio in each pixel, which we calculate according to the following formula:

$$R_p' = \sum_q \widetilde{m}_{s,p,q} R_q. \tag{17}$$

However, for numerical reasons, minimization in the general form given in Equation (16) with this substitution would be too complicated. In our minimization scheme, we instead use an iterative procedure, described below, to obtain subsequent estimations of the compression ratio vector, and we use the result from the previous iteration in Equation (17) in the next iteration. The Compton-Getting corrections are calculated for compression ratios 2.0, 2.5, 3.0, 3.5, and 4.0. Between these values, we use linear interpolation to get the value of $c_{\tau,p,s}(R'_p)$. We denote the values at the calculated compression ratios as $c_{\tau,p,s,r}$.

As the model flux is calculated for a finite number of energy bins and compression ratio values, we use a two-step interpolation scheme. The model is calculated for 66 logarithmically spaced energies $e$ from 0.1 keV to 15 keV and for the same compression ratio values as the Compton-Getting correction. Let $\tilde{g}_{\tau,q,s,e,r}$ denote the model value for calculated energy $e$ and compression ratio $r$. Because, for each HTS position, the energy grid is the same, we find the two nearest logarithms of the energy bins $\log e_1$ and $\log e_2$ to $\log(c_{\tau,p,s}(R'_p)\bar{E}_{\tau,p,s})$. With these two, we define the following transformation tensor:

$$\mathbf{T}_{\tau,p,s,e,r} = \begin{cases} \frac{\log e_1 - \log(c_{\tau,p,s}(R'_p)\bar{E}_{\tau,p,s})}{\log e_1 - \log e_2} & \text{if } e = e_2 \\ \frac{\log e_2 - \log(c_{\tau,p,s}(R'_p)\bar{E}_{\tau,p,s})}{\log e_2 - \log e_1} & \text{if } e = e_1 \\ 0 & \text{otherwise} \end{cases}. \quad (18)$$

Using this tensor, we interpolate over the compression ratio using the following formula:

$$g_{\tau,q,s}(c_{\tau,p,s}(R'_p)\bar{E}_{\tau,p,s}, R_q) = \sum_e \mathbf{T}_{\tau,p,s,e,r_q^0} \tilde{g}_{\tau,q,s,e,r_q^0} + $$
$$(R_q - r_q^0) \frac{\sum_e \mathbf{T}_{\tau,p,s,e,r_q^+} \tilde{g}_{\tau,q,s,e,r_q^+} - \sum_e \mathbf{T}_{\tau,p,s,e,r_q^-} \tilde{g}_{\tau,q,s,e,r_q^-}}{r_q^+ - r_q^-}. \quad (19)$$

In this last equation $r_q^0$, $r_q^+$, $r_q^-$ denote the closest, closest among larger, and closest among smaller values ($r_q^0 = r_q^+$ or $r_q^0 = r_q^-$) from the above set to the $R_q$ value obtained from the previous iteration within available compression ratios.

The minimization of Equation (16) is, in most situations, ill-conditioned because of the high number of fit parameters. Consequently, regularization is needed to minimize Equation (16). We use the Tikhonov regularization method[101,102] with the following regularization term,

$$\chi^2_{\text{reg}} = \sum_q \sum_{q' \in N(q)} \frac{(R_q - R_{q'})^2}{\text{dist}(q,q')^2}, \quad (20)$$

where $N(q)$ gives 8 pixels around pixel $q$, and $\text{dist}(q, q')$ is the angular distance between the centers of a pair of pixels. We limit the sum to the nearest neighbors for computational reasons.

In our fitting, we minimize the following sum,

$$\chi^2(\lambda) = \chi^2_{\text{LS},\tau} + \lambda \chi^2_{\text{reg}}, \quad (21)$$

where $\lambda$ is the regularization parameter. We find the optimal regularization parameter using the L-curve technique[103]. Namely, we minimize Equation (21) as a function of $\lambda$, and we inspect the trajectory of the optimal $(\log \chi^2_{\text{LS},\tau}, \log \chi^2_{\text{reg}})$. The curve made by this trajectory has an L-shape (Figures S11a and S11c), and we seek the corner where the curvature of the trajectory is the largest (Figures S11b and S11d). The solution found in the corner is adopted as our best fit. Note that while $\chi^2_{\text{LS},\tau}$ does not vary as significantly as $\chi^2_{\text{reg}}$, compression ratio maps produced from the right-most edges of the L-curve are overly and unrealistically smoothed due to a large regularization parameter.

The interpolations discussed previously are initially spanned between compression ratios 2.5 and 3.0 for each point at the HTS. After obtaining the best fit, we span the interpolations in subsequent iterations between the two nearest values to the compression ratio determined in the previous iteration separately for each direction. We then continue to iterate until the problem converges. While we find there is a small set of pixels in which the compression ratio is close to one of the values used for the modeling, and the best-fit values oscillate between values slightly above and below this value (e.g., slightly above 3.0 and then slightly below 3.0), the range of these oscillations is much smaller than the uncertainties derived as described below.

To calculate the uncertainty of the derived compression ratio, we start with the uncertainty related to the IBEX data uncertainty. This uncertainty can be calculated from the inverted matrix of second derivatives around the minimum of Equation (21) calculated for the optimal $\lambda_{opt}$:

$$\mathbf{V} = \frac{\chi^2_{\text{LS},\tau,\text{min}}}{n_{\text{dof}}} \left( \frac{1}{2} \frac{\partial^2 \chi^2_{\text{min}}(\lambda_{opt})}{\partial \mathbf{R} \partial \mathbf{R}} \right)^{-1}. \tag{22}$$

Square roots of the diagonal terms of this matrix give the uncertainties for each pixel and time period, $\sigma_{\tau,p}$. The uncertainty matrix is multiplied by the reduced $\chi^2_{\text{LS}}$. The number of degrees of freedom is generally given as [(# ESA steps used in the fit) – 1] × (# pixels with the data) – 1800.

Next, we check how the fit compression ratios respond to uncertainties in the (1) SW speed at 1 au, (2) density of interstellar neutral hydrogen in the heliosphere, (3) SW-to-ENA time delay, (4) HTS microstructure used in our test particle model, (5) distance to the HTS, (6) core SW temperature at 1 au, (7) the diffusion coefficient spectral index, and (8) the initial kappa index of the proton distribution just downstream of the HTS. For this, the calculations are repeated with modified modeled proton fluxes and Compton-Getting correction factors corresponding to these changes, where we add or subtract 1-σ values to each variable. The 1-σ uncertainties are assumed to be: (1) –5% for SW speed[52], where we choose to subtract uncertainties to mimic the shifting of IPS-derived SW speeds to match OMNI data at low latitudes (during these time periods, the IPS-derived SW speeds tend to overestimate the OMNI data). We note that ~10% of the IPS-derived SW speeds during these two time periods is the maximum shift required to match in-situ observations of the SW near the ecliptic plane (Figure 11 in Porowski et al.[52]). Thus, we propagate –10% uncertainties in the SW speed through our analysis, then take half of their summed differences from the nominal case to approximate the 1-σ uncertainty (i.e., half of 10%); (2) –10% for interstellar neutral H density[85], where we subtract by 1-σ due to the historical value of interstellar H density[104] being lower than the currently-accepted value[85]; (3) +33% for SW-to-ENA time delay as a rough estimate for uncertainty (no preference for adding or subtracting 1-σ), to account for uncertainties related to using a steady-state MHD simulation's flow streamlines in the HS; (4) changing the HTS foot width to 1.5 $L_R$, based on uncertainties in the Voyager 2 shock speed uncertainty[6,34], as our estimate of 1-σ uncertainty; (5) –5% for the distance to the HTS, $l_{HTS}$, where we choose to subtract 1-σ to mimic the effects of increasing the SW speed (i.e., a closer HTS means less time slowed down by mass-loading). We choose a ±5% uncertainty which is approximately ±5 au in the nose-ward hemisphere, based on how the HTS may move over time informed by dynamic models[105] and observations[15]; (6) +10% for the core SW temperature at 1 au, particularly because of the lack of data at high latitudes without Ulysses. We choose 10% because it approximately reflects the variation in high-latitude SW speed variations[58,59]; (7) ±0.2 for the diffusion coefficient spectral index, where we take half of the summed differences from the nominal case as the uncertainty for $\alpha$; and (8) ±0.2 for the kappa index, chosen based on the fact that the majority of data lie between kappa ~ 2 and 2.4 (with kappa ~ 2 close to the mode of the all-sky distribution, and highly skewed to larger values) based on analyses of IBEX data[26] and their agreement with test particle results[6,7]. The uncertainty is also calculated using half of the summed differences from the nominal case, i.e., taking their average.

We find the final, best-fit compression ratio per time period and pixel, $R^f_{\tau,p}$, shown in Figures 5a and 5b. The total uncertainty of the compression ratio is given as:

$$\sigma_{tot,\tau,p} = \{\sigma^2_{\tau,p} + \frac{1}{4}\left[R^f_{\tau,p} - R_{\tau,p}(u - \sigma_{u,10\%})\right]^2$$

$$+\left[R^f_{\tau,p} - R_{\tau,p}(n_H + \sigma_n)\right]^2 + \left[R^f_{\tau,p} - R_{\tau,p}(t + \sigma_t)\right]^2$$

$$+\left[R^f_{\tau,p} - R_{\tau,p}(L_{HTS} + \sigma_L)\right]^2 + [R^f_{\tau,p} - R_{\tau,p}(d_{HTS} - \sigma_d)]^2$$

$$+\left[R^f_{\tau,p} - R_{\tau,p}(T + \sigma_T)\right]^2$$

$$+ \left[\frac{1}{2}\left(\left|R^f_{\tau,p} - R_{\tau,p}(\alpha + \sigma_\alpha)\right| + \left|R^f_{\tau,p} - R_{\tau,p}(\alpha - \sigma_\alpha)\right|\right)\right]^2$$

$$+ \left[\frac{1}{2}\left(\left|R^f_{\tau,p} - R_{\tau,p}(\kappa + \sigma_\kappa)\right| + \left|R^f_{\tau,p} - R_{\tau,p}(\kappa - \sigma_\kappa)\right|\right)\right]^2\}^{1/2}, \qquad (23)$$

where the first term is the propagated statistical uncertainty of $R^f_{\tau,p}$. The second term contains a factor of 1/4 which combines the averaging and halving to obtain the desired (5%) 1-σ uncertainty. Note that the terms in parentheses in Equation (23) represent parameters of $R_{\tau,p}$. The maps of the total uncertainties are shown in Figures 5c and 5d.

**Figures/Tables**

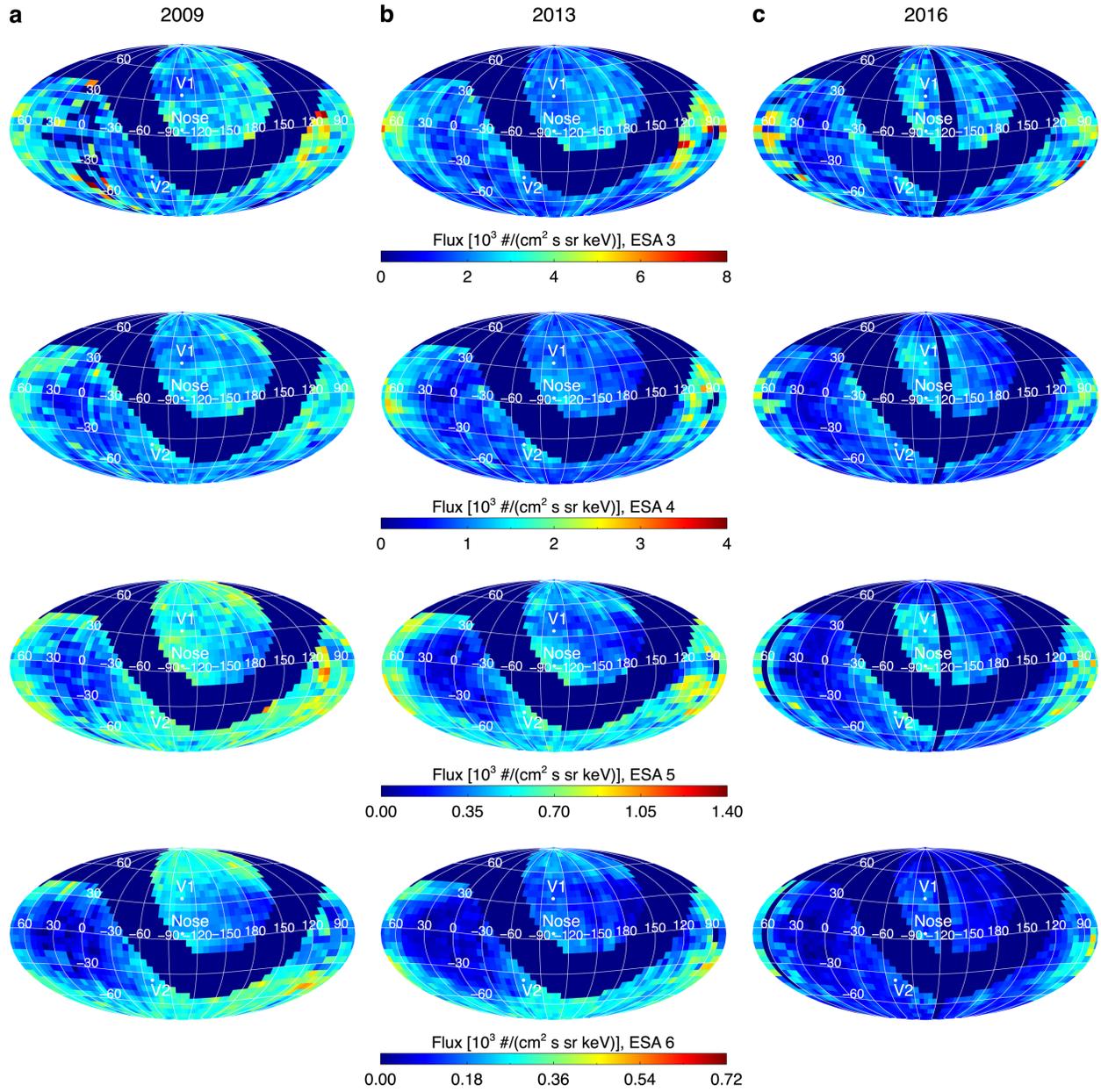

*Figure 1.* Example sky maps of proton fluxes in the HS plasma frame derived from IBEX-Hi ENA observations [51]. We show maps in (a) 2009, (b) 2013, and (c) 2016 for ESA 3-6. Note that pixels near the ribbon are removed. Also note that the proton energy per pixel is different due to the Compton-Getting correction.

**Upstream SW Speed
Corresponding to IBEX Timing**

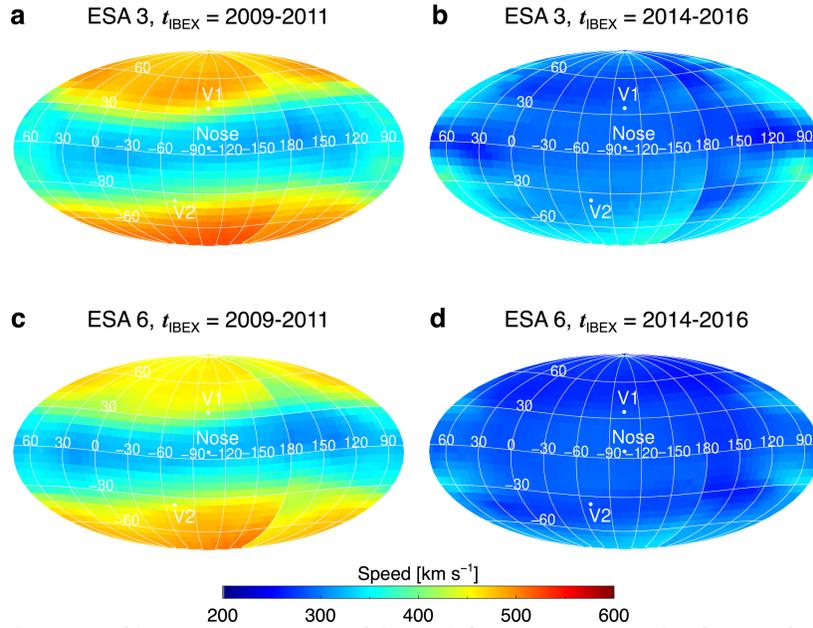

*Figure 2.* Example sky maps of SW speed upstream of the HTS for two time periods of IBEX observations (2009-2011 and 2014-2016). The maps corresponding to ESA 3 and 6 demonstrate how the ENA time delays are different for each ESA (see also Figure S3). IBEX's Sun-pointing spinning and the time it takes to observe the entire sky are included.

**PIC Simulation Results**

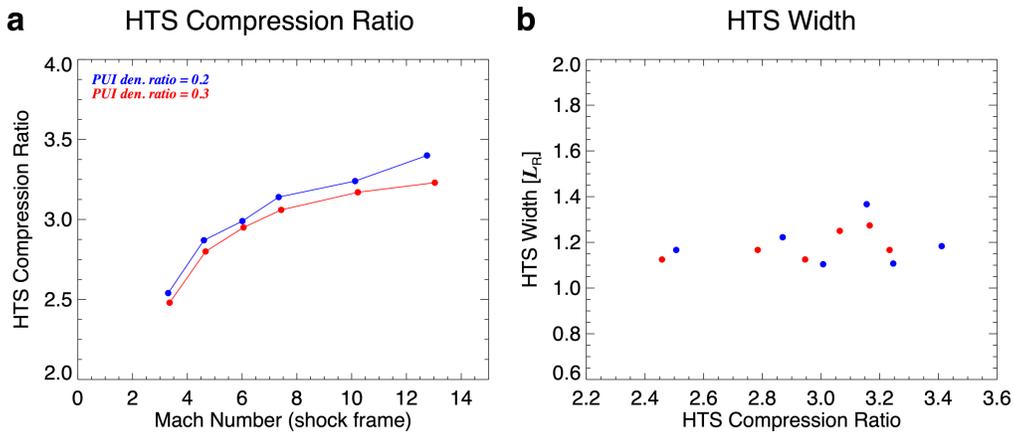

*Figure 3.* (a) PIC simulation results for different upstream Alfvén Mach numbers and PUI density ratios (PUI to total proton density). The resulting compression ratio of the shock depends strongly on the upstream Mach number, but weakly on the PUI density ratio. (b) PUI foot + ramp width ("shock width") as a function of shock compression ratio. The shock width is weakly dependent on the compression ratio (and upstream Mach number). Note that the runs in panel b are the same as those in panel a, color-coded by their respective PUI density ratio.

**Downstream PUI Distributions from Test Particle Simulation**

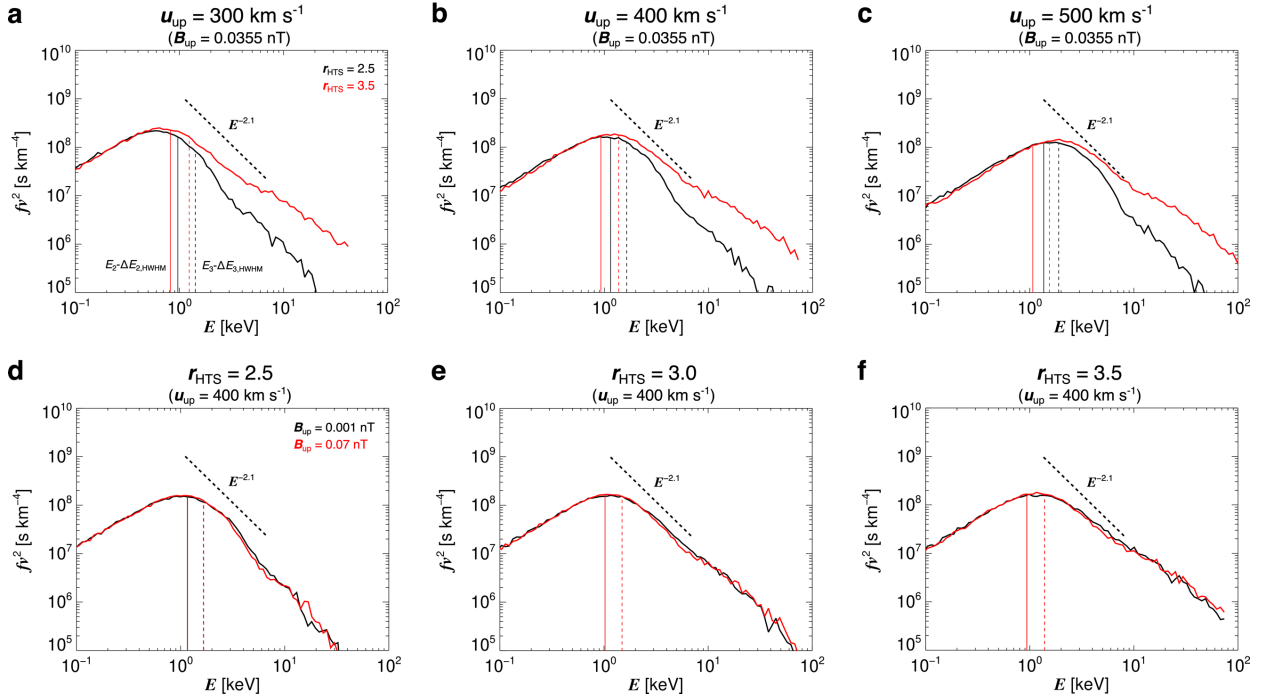

*Figure 4.* Examples of downstream PUI distributions from our test particle simulation for different upstream conditions (using case 1 width of ~1 $L_R$). In panels a-c we show results for different HTS compression ratios ($r_{HTS}$ = 2.5, 3.5) as a function of upstream SW speed ($u_{up}$ = 300, 400, 500 km s$^{-1}$). In panels d-f we show results for different upstream magnetic field ($B_{up}$ = 0.001, 0.07 nT) as a function of HTS compression ratio ($r_{HTS}$ = 2.5, 3, 3.5). The vertical solid lines show the central energy of ESA 2 minus ESA 2's half width at half maximum. The vertical dashed lines show the same, except for ESA 3. Also shown is a power law with slope of $E^{-2.1}$ for reference[51]. Note that the 2 vertical solid lines and 2 vertical dashed lines in panels d-f overlap each other.

# Sky Maps and Surface Plots of the HTS Compression Ratio

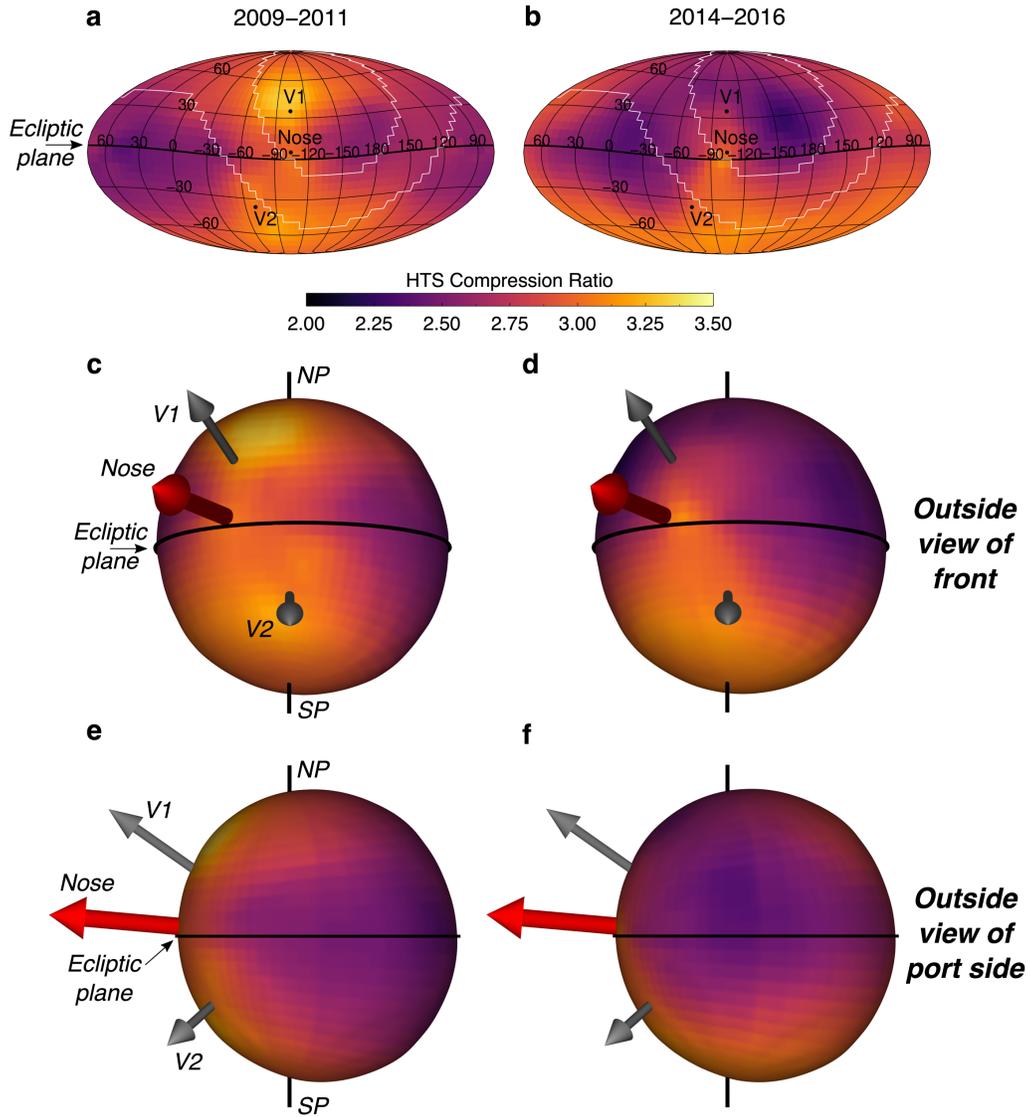

*Figure 5.* (a, b) Sky maps of the HTS compression ratio for IBEX time periods (a) 2009-2011 and (b) 2014-2016. (c, d) 3D surface plots of the compression ratio from two viewpoints. The size of the HTS is derived from our MHD simulation, and the color is the HTS compression ratio value. Panels c and d show views of the front of the heliosphere

*(offset from the Nose), and panels e and f show views of the port side. We show arrows for the Nose, V1, and V2 trajectories, and lines for the ecliptic plane, the ecliptic north pole (NP), and ecliptic south pole (SP). (g, h) Sky maps of the HTS compression ratio uncertainty for both time periods. The white contour in the sky maps shows the region of data surrounding the ribbon excluded from our data input to the minimization process. The high values of the relative uncertainties near the port side of the heliosphere (~15° longitude and ~200° longitude in 2014-2016) are connected to the regularization uncertainty, which result in different levels of smoothness near the local minimum (in compression ratio) for considered variations of the SW parameters. The higher uncertainty in 2009-2011 from the polar regions is due to the increase in uncertainty from the fast SW speed. The higher uncertainties near the port and starboard lobes in 2014-2016 are due to the regularization minimization procedure attempting to find the best compression ratio in the spatially small, low compression regions. Smoother transitions in compression across the sky, such as in 2009-2011, yield lower uncertainties after minimization.*

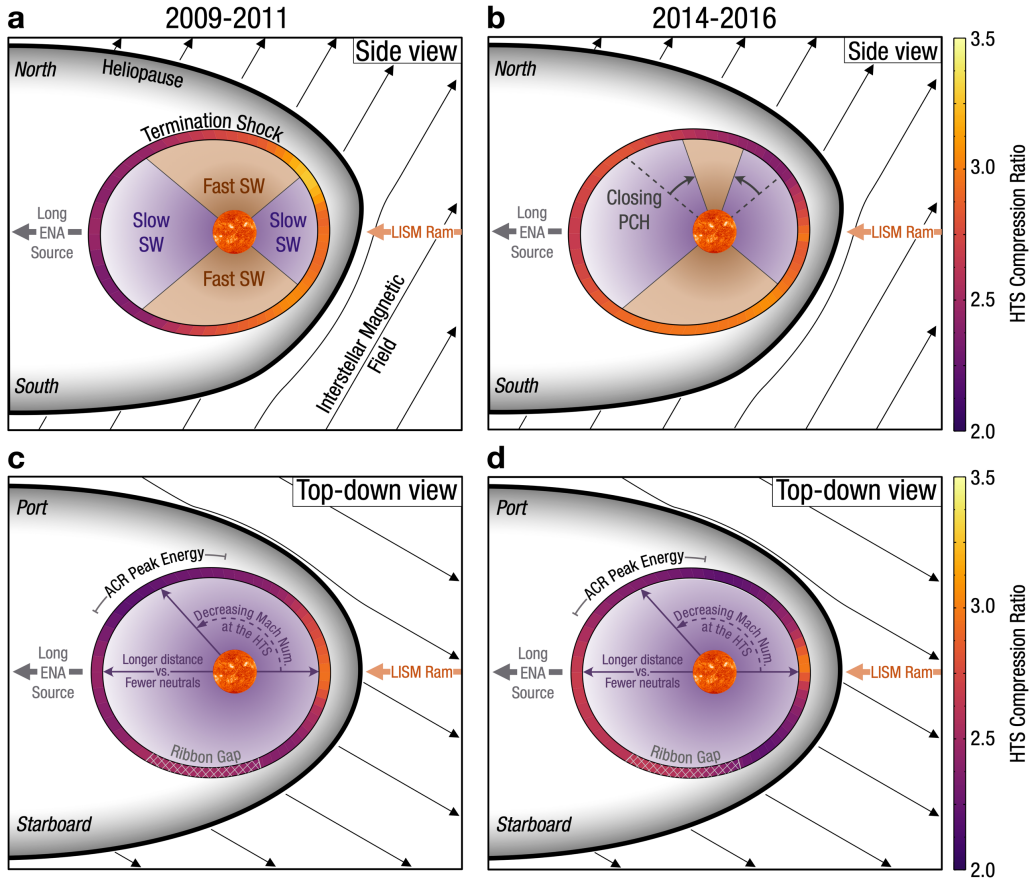

*Figure 6. Illustrations of the HTS compression ratio in the solar meridional ("side view", panels a,b) and solar equatorial ("top-down view", panels c,d) cross sections through the heliosphere, and the physical mechanisms responsible for the measured shock compression. The surface of the HTS is color-coded with the respective HTS compression ratio, for the different time periods (panels a,c, 2009-2011; panels b,d, 2014-2016). Regions and boundaries of the heliosphere are labeled, and the fast vs. slow SW are color-coded as tan and purple, respectively. The proton density as a function of distance from the Sun is illustrated by the gradients in these colors. The five primary variables controlling the observations are (1) LISM ram pressure, (2) SW speed, (3) closing of the PCH, and (4) the amount of SW mass-loading with distance to the HTS. Note that the shapes of the heliosphere boundaries in the left panels are adapted from McComas et al.[106].*

*Table 1 | HTS compression ratio in selected directions of the sky.*

| Direction [a] | 2009-2011 | | 2014-2016 | | Statistically significant change? [e] | Physical causes for observed HTS compression relative to other locations or times |
| --- | --- | --- | --- | --- | --- | --- |
| | $r_{HTS}$ [b] | $\sigma_r$ | $r_{HTS}$ | $\sigma_r$ | | |
| Nose [b] | 2.96 | 0.24 | 2.85 | 0.22 | No | LISM pressure |
| Voyager 1 | 3.25 | 0.24 | 2.70 | 0.14 | Yes | PCH-SW properties/evolution |
| Voyager 2 [c] | 3.03 | 0.26 | 2.96 | 0.14 | No | PCH-SW properties |
| Port lobe [d] | 2.48 | 0.16 | 2.51 | 0.17 | No | SW slowing due to distance to HTS |
| Starboard lobe [d] | 2.59 | 0.16 | 2.47 | 0.13 | Maybe | SW slowing due to distance to HTS |
| North pole | 2.90 | 0.29 | 2.63 | 0.10 | Maybe | PCH-SW evolution, ISMF |
| South pole | 2.97 | 0.33 | 3.06 | 0.14 | No | PCH-SW evolution, ISMF |
| Central Tail | 2.61 | 0.14 | 2.79 | 0.16 | Maybe | SW slowing due to distance to HTS vs. less mass-loading due to fewer interstellar neutrals |

**Notes.**

[a] Directions in ecliptic J2000 (longitude, latitude): Nose = (255.7°, 5.1°) (however, see note b below), Voyager 1 = (256.1°, 35.1°), Voyager 2 = (290.3°, -36.4°), Port lobe = (14.6°, -8.4°), Starboard lobe = (162.6°, 24.9°), North pole = (0°, 90°), South pole = (0°, -90°), Central Tail = (75.7°, -5.1°).

[b] Compression ratios and their uncertainties are taken from a single 6°x6° IBEX pixel nearest to the desired direction. However, due to the higher systematic uncertainties of the regularization routine in smoothing around narrow regions of peaks (i.e., near the Nose) of the compression ratio, particularly in 2014-2016, we take the Nose pixel to be centered at (249°, 9°), shifted one pixel northward and one pixel starboard. This also moves it away from the ribbon mask region.

[c] Voyager 2 observations measured a large-scale HTS compression between ~2.5 and 3.0, depending on the scale over which the plasma properties are used to calculate the compression, i.e., sub-shock vs. energetic particle precursor scales[36,37,63]. It is interesting to note that the simple theoretical formulation for PUI heating across the shock from Shrestha et al.[49] provides a compression ratio closer to the small-scale compression ratio of 2.5, although without providing uncertainties. Note that Voyager 2 crossed the HTS multiple times, with compression ratio values at micro-scale sizes of 2.38 ± 0.14 and 1.58 ± 0.71 at TS-2 and TS-3 crossings, respectively[35], or an average of 1.98 ± 0.36. Our analysis, however, only focuses on the large-scale compression.

[d] We take the port and starboard lobe directions to be the lobe centers found by Dayeh et al.[61]. To avoid confusion over the change in time of the lobe centers, we take the arithmetic mean of the centers found by Dayeh et al. in 2009-2011 and 2014-2015, yielding (14.6°, -8.4°) and (162.6°, 24.9°) for the port and starboard lobe centers, respectively.

[e] Significant changes are determined to be "yes" if the compression ratio values observed in 2009-2011 and 2014-2016 lie completely outside their counterpart 1-σ uncertainty range though there may be overlap of the uncertainty ranges; "no" if the compression ratio values lie close to each other within both uncertainty ranges; and "maybe" if the compression ratios lie just inside their counterpart 1-σ uncertainty range.

**Acknowledgements**

This work was supported by NASA's Heliophysics Supporting Research program under grant no. 80NSSC21K1686. E.J.Z. acknowledges partial support from Laboratory Directed Research and Development (LDRD) program of Los Alamos National Laboratory (LANL) under project number 20220107DR, subcontract number CW18805. E.J.Z., B.S., and J.R.S. acknowledge partial support from NASA's IBEX Mission (80NSSC20K0719), which is part of the Explorers Program. P.S. acknowledges partial support from the Polish National Agency for Academic Exchange within the Polish Returns Programme (BPN/PPO/2022/1/00017) and the National Science Centre, Poland (2023/02/1/ST9/00004). E.J.Z. thanks Czeslaw Porowski for providing tabulated SW speed and density model results based on IPS and ACE/Wind observations. E.J.Z also thanks Joe Giacalone, Dave McComas, and Gary Zank for helpful discussions related to the methodology and results of this study.

# Supplementary Information

### for

### Global Heliospheric Termination Shock Strength from the Solar-Interstellar Interaction

E. J. Zirnstein[1], R. Kumar[1,2], B. L. Shrestha[1], P. Swaczyna[1,3], M. A. Dayeh[4,5], J. Heerikhuisen[6], J. R. Szalay[1]

[1]Department of Astrophysical Sciences, Princeton University, Princeton, NJ 08544, USA
(ejz@princeton.edu)

[2]Tau Systems, Austin, TX 78701, USA

[3]Space Research Centre PAS (CBK PAN), Bartycka 18a, 00-716, Warsaw, Poland

[4]Southwest Research Institute, San Antonio, TX 78238, USA

[5]Department of Physics and Astronomy, University of Texas at San Antonio, San Antonio, TX 78249, USA

[6]Department of Mathematics and Statistics, University of Waikato, Hamilton, New Zealand

Here we provide supplemental figures referenced in the main text and Methods section of the paper.

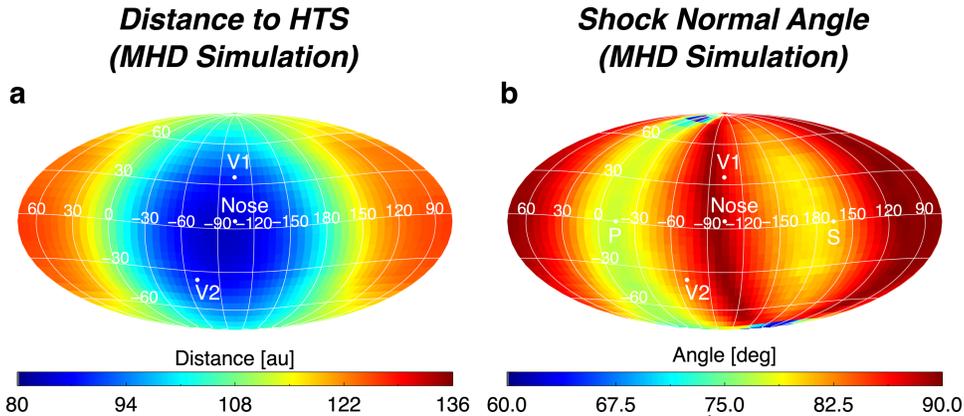

**Figure S1.** *(a) Distance to the HTS from the MHD simulation used in this study[1]. Distances were scaled such that the average distance to the HTS from the simulation in the V1 and V2 directions is the same as the average of the observed crossings from the Voyager spacecraft[2,3]. (b) Shock normal angle from the MHD simulation. For most of the sky, the shock normal angle is ≳75°, except very close to the solar heliographic poles (tilted by 7.25° from the ecliptic poles). Both maps are plotted in ecliptic J2000 coordinates.*

**SW Plasma Properties Upstream of HTS**

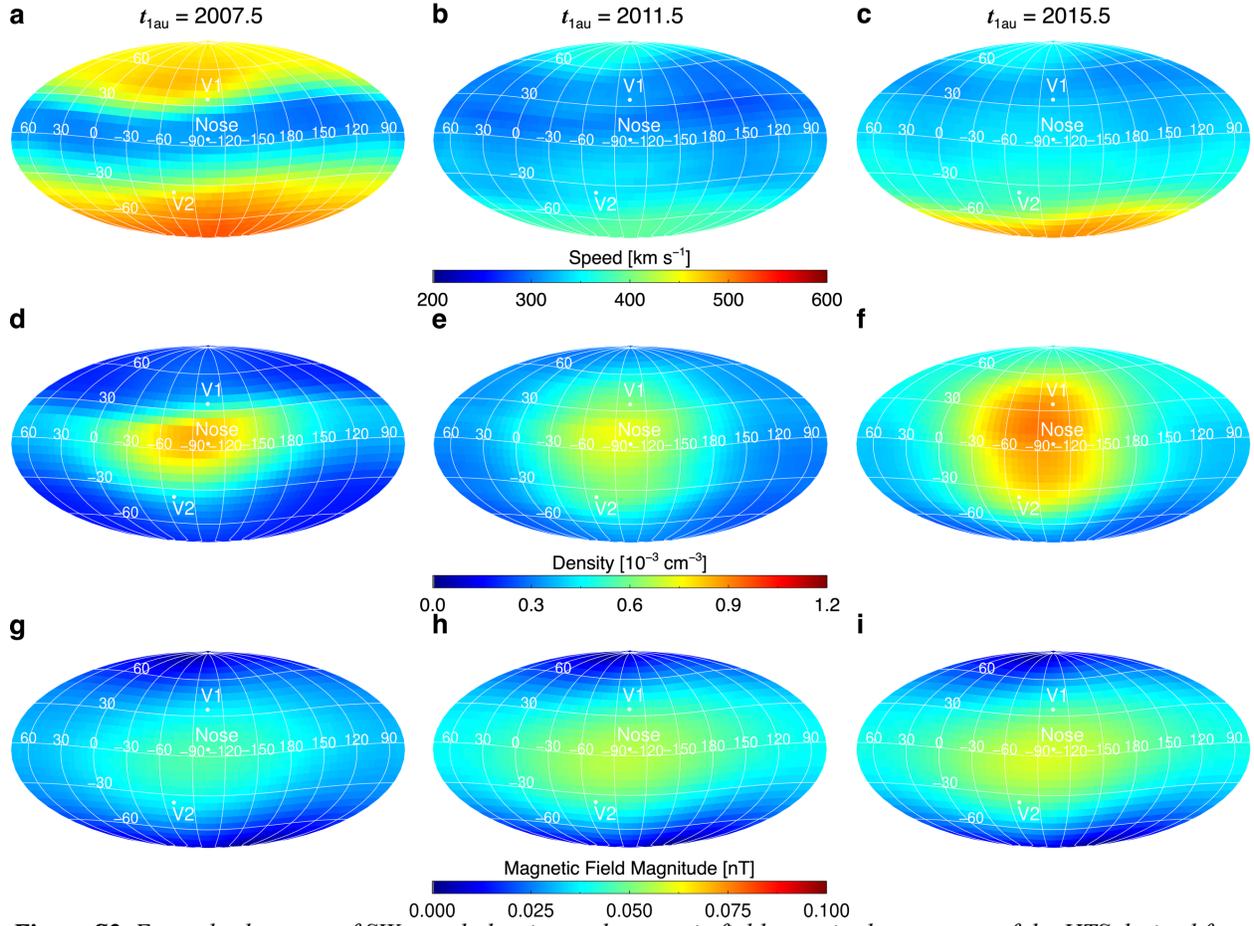

*Figure S2.* Example sky maps of SW speed, density, and magnetic field magnitude upstream of the HTS derived from IPS/OMNI data taken at 1 au in (a) 2007.5, (b) 2011.5, and (c) 2015.5 and propagated to the HTS using Eq. (1)-(5). Note that the results for $t_{1au}$ = 2015.5 reflect the enhancement in SW pressure beginning at the Sun in late 2014, but we do not use this data in our analysis. We restrict our analysis to times before IBEX first observed an increase in ENA fluxes associated with the SW pressure increase.

**Total SW-to-ENA Time Delay**

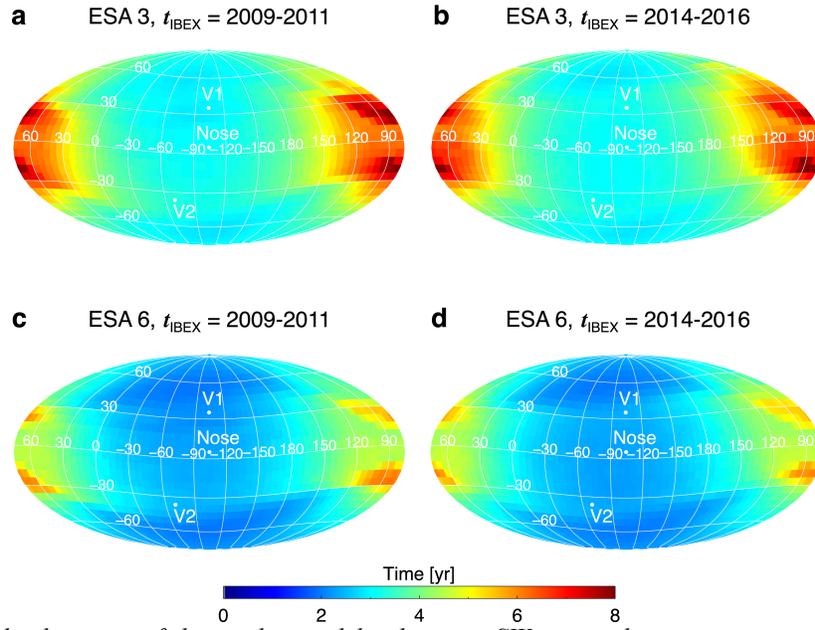

*Figure S3.* Example sky maps of the total time delay between SW outward propagation at 1 au and IBEX ENA observation, for ESA 3 (a, b) and ESA 6 (c, d). These results are based on flow streamlines from a steady-state simulation of the heliosphere[1], but speeds based on the IPS/OMNI model[4]. See text for details.

**Upstream SW Magnetic Field Corresponding to IBEX Timing**

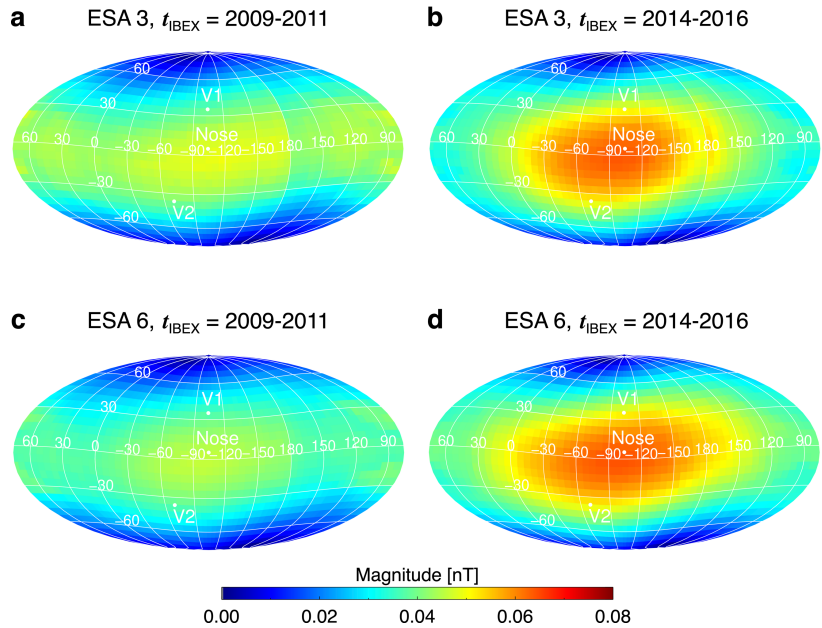

*Figure S4.* Similar to Figure 2 in the main text, except we show magnetic field magnitude.

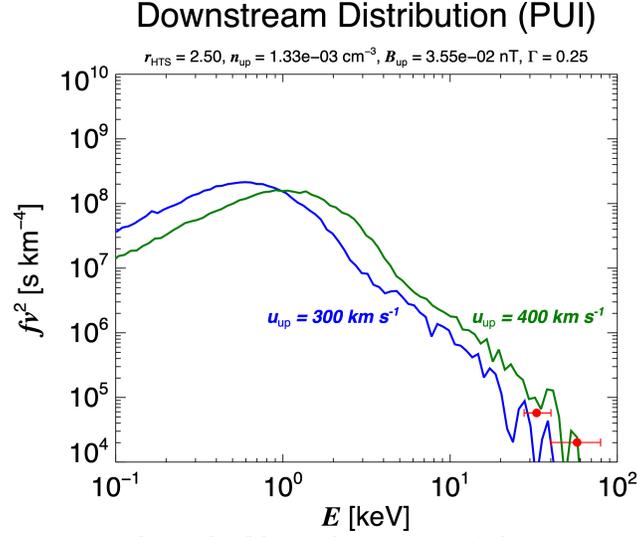

***Figure S5.*** *Comparison of our test particle results (blue and green curves) downstream of the HTS to Voyager 2 LECP data[5,6] (red dots with error bars). Note that the test particle model has low statistics at energies above ~15 keV, but the model spectra still reproduce the LECP observations. Here, the PUI density ratio is $\Gamma = 0.25$, same as in Figure 4 in the main text.*

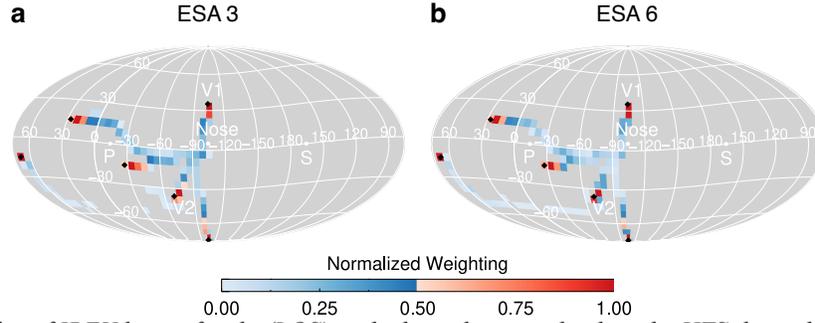

***Figure S6.*** *Examples of IBEX lines of sight (LOS) and where they map back to the HTS depending on the HS plasma flow streamlines. Examples are shown for ENAs at energies for ESA 3 (a) and ESA 6 (b) are shown. The colored pixels are positions on the HTS that are connected to the IBEX LOS (black diamond symbols) via bulk plasma flow streamlines in the global MHD simulation. Note that the weights for ESA 3 vs. 6 are different due to the energy-dependent source regions of the ENAs. Because of this, several pixels with weights close to zero appear different in panels a and b, but the flow streamlines themselves are the same.*

**Proton Intensity Scaling Factor**

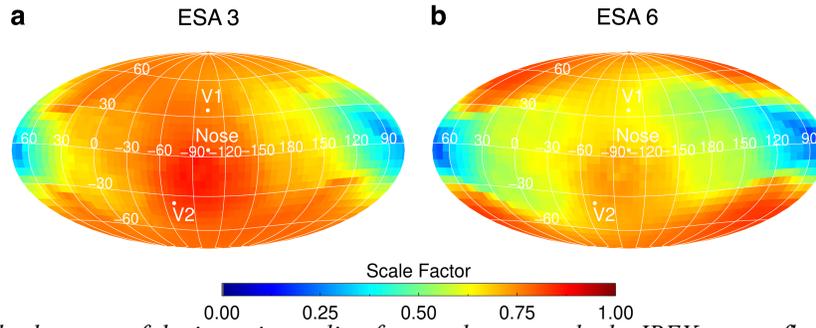

*Figure S7. Example sky maps of the intensity scaling factors that we scale the IBEX proton fluxes with as a function of ESA passband, in order to "undo" the evolution of the proton distribution through the HS.*

**Comparison of Simulated and Observed GDF Maps**

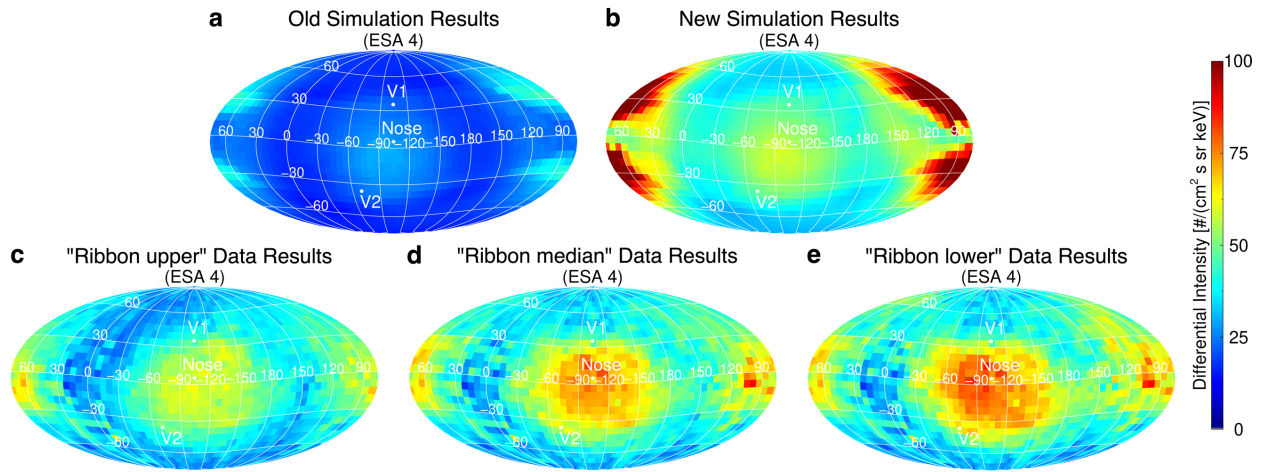

*Figure S8. Examples of simulated and IBEX observation-based maps of the GDF in 2009-2011 for ESA 4 (~1.7 keV). The ENA simulation results are shown from two methods: first from the methodology in Zirnstein et al.[1] (panel a) and from the methods described in this study (panel b). The observation-based maps are shown in panels c-e. Note that the inclusion of the velocity diffusion effect increased the flux globally such that the simulation results compare much better to the data than in some previous studies[7–9]. The oversaturation of simulated fluxes from the mid-latitude, north/south tailward directions is not surprising, as they are due to the fast SW propagating down the heliotail in our steady-state MHD simulation, without the cyclic transitions from fast to slow to fast, etc., SW. The observational data are from IBEX Data Release 18 as validated in McComas et al.[10], based on the ribbon separation methods by Beesley et al.[11]*

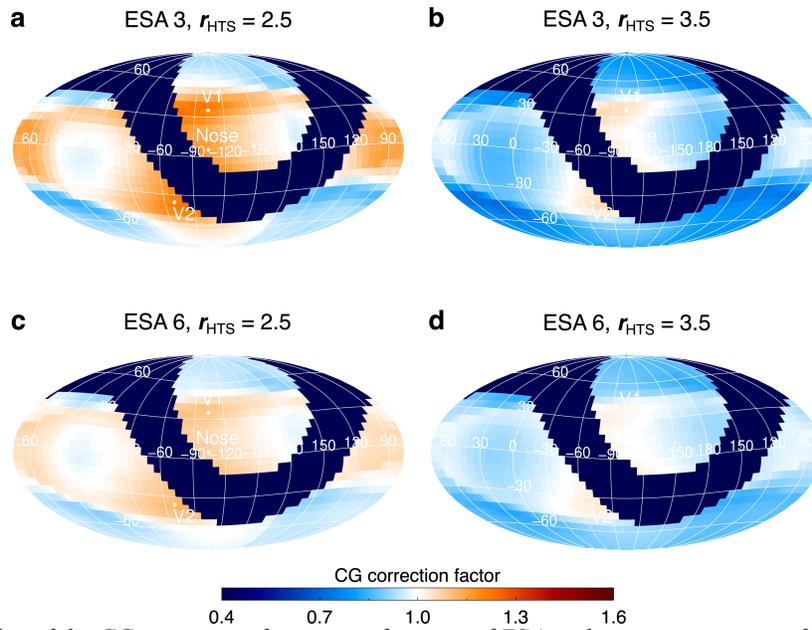

*Figure S9.* Examples of the CG correction factor as a function of ESA and compression ratio for 2009-2011.

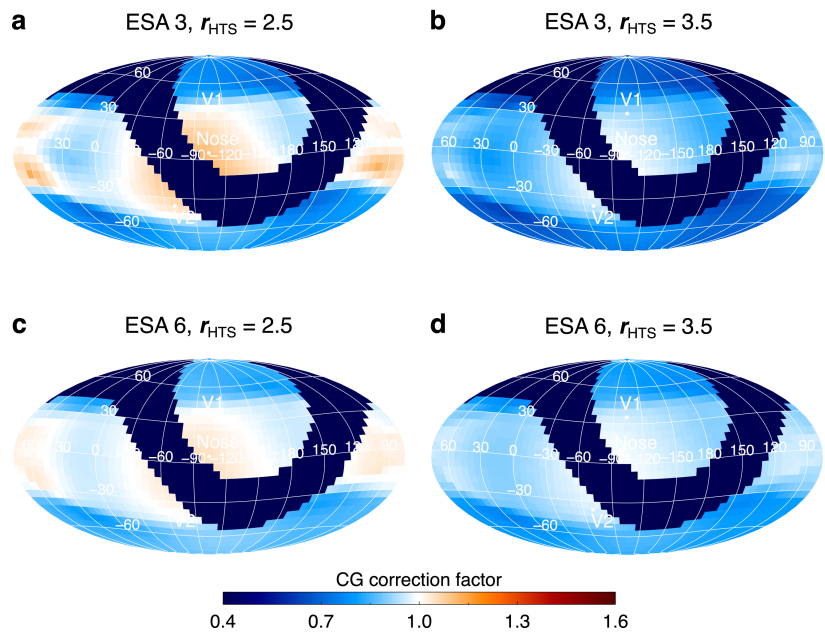

*Figure S10.* Similar to Figure 9, except for 2014-2016.

**Least-squares and Regularization Minimization**

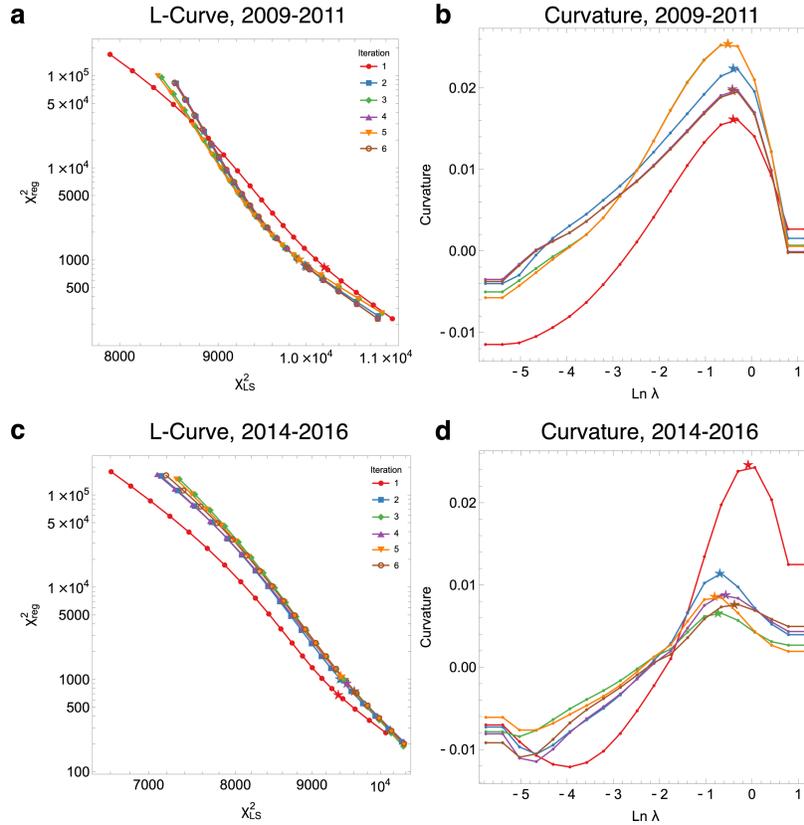

*Figure S11.* Example of the L-curve (a,c) and its curvature (b,d) used to determine the best-fit compression ratio simultaneously for all pixels in a map. These results are from the nominal case (L1), shown for time periods 2009-2011 (a,b) and 2014-2016 (c,d). The point of maximum curvature is shown as the star in both panels. Note that the stars in panels b and d do not necessarily lie on the curve because the point of maximum curvature is found by fitting a Gaussian function to the 5 points nearest to the peak.